%% file: astroph.tex
\documentclass[apj]{emulateapj}
\usepackage{float,epsfig}
\usepackage{pstricks}

\newcommand{\etal}{\mbox{et~al.}}

\def\deg      {{\ifmmode^\circ\else$^\circ$\fi}} 

 \shorttitle{Clusters in the COSMOS field}
 \shortauthors{Finoguenov et al.}
 
\begin{document}

\submitted{to appear in COSMOS ApJS special issue 2007} 
\title{The XMM-Newton wide-field survey in the COSMOS field:
VI. Statistical properties of clusters of galaxies.\altaffilmark{$\star$}}

\altaffiltext{ }{$\star$ Based on observations with the XMM-Newton, an ESA
  science mission with instruments and contributions directly funded by ESA
  Member States and NASA; the NASA/ESA {\em Hubble Space Telescope},
  obtained at the Space Telescope Science Institute, which is operated by
  AURA Inc, under NASA contract NAS 5-26555; also based on data collected at
  : the Subaru Telescope, which is operated by the National Astronomical
  Observatory of Japan; the European Southern Observatory, Chile; Kitt Peak
  National Observatory, Cerro Tololo Inter-American Observatory, and the
  National Optical Astronomy Observatory, which are operated by the
  Association of Universities for Research in Astronomy, Inc. (AURA) under
  cooperative agreement with the National Science Foundation; the National
  Radio Astronomy Observatory which is a facility of the National Science
  Foundation operated under cooperative agreement by Associated
  Universities, Inc ; and the Canada-France-Hawaii Telescope operated by the
  National Research Council of Canada, the Centre National de la Recherche
  Scientifique de France and the University of Hawaii.}

\author{A. Finoguenov\altaffilmark{1,2}, L. Guzzo\altaffilmark{3}, G. Hasinger\altaffilmark{1},
N. Z. Scoville\altaffilmark{4,5}, H. Aussel\altaffilmark{6}, H. B\"ohringer\altaffilmark{1},
M. Brusa\altaffilmark{1}, P. Capak\altaffilmark{4}, N. Cappelluti\altaffilmark{1}, A. Comastri\altaffilmark{7}, S. Giodini\altaffilmark{3},
R. E. Griffiths\altaffilmark{13}, C. Impey\altaffilmark{8}, A. M. Koekemoer\altaffilmark{9},
J.-P. Kneib\altaffilmark{10}, A.  Leauthaud\altaffilmark{10}, O. Le F\`evre\altaffilmark{10},
S. Lilly\altaffilmark{12}, V. Mainieri\altaffilmark{1,19}, R. Massey\altaffilmark{4},
H. J. McCracken\altaffilmark{15,16}, B. Mobasher\altaffilmark{9}, T. Murayama\altaffilmark{17},
J. A. Peacock\altaffilmark{14}, I. Sakelliou\altaffilmark{11}, E. Schinnerer\altaffilmark{11},
J. D. Silverman\altaffilmark{1}, V. Smol{\v c}i{\' c}\altaffilmark{11}, Y. Taniguchi\altaffilmark{18},
L. Tasca\altaffilmark{10}, J. E. Taylor\altaffilmark{4}, J. R. Trump\altaffilmark{8}, G. Zamorani\altaffilmark{7}}

\altaffiltext{1}{Max-Planck-Institut f\"ur Extraterrestrische Physik,
             Giessenbachstra\ss e, 85748 Garching, Germany}
\altaffiltext{2}{University of Maryland, Baltimore County, 1000
  Hilltop Circle,  Baltimore, MD 21250, USA}
\altaffiltext{3}{Osservatorio Astronomico di Brera, via Bianchi 46,
  23807 Merate, Italy}
\altaffiltext{4}{California Institute of Technology, MC 105-24, 1200 East
California Boulevard, Pasadena, CA 91125, USA}
\altaffiltext{5}{Visiting Astronomer, Univ. Hawaii, 2680 Woodlawn Dr.,
  Honolulu, HI, 96822, USA}
\altaffiltext{6}{Service d'Astrophysique, CEA/Saclay, 91191 Gif-sur-Yvette, France}
\altaffiltext{7}{INAF-Osservatorio Astronomico di Bologna, via Ranzani 1, 40127
  Bologna, Italy}
\altaffiltext{8}{Steward Observatory, University of Arizona, 933 North Cherry Avenue,
  Tucson, AZ 85721, USA}
\altaffiltext{9}{Space Telescope Science Institute, 3700 San Martin
Drive, Baltimore, MD 21218, USA}
\altaffiltext{10}{Laboratoire d'Astrophysique de Marseille, BP 8, Traverse
du Siphon, 13376 Marseille Cedex 12, France}
\altaffiltext{11}{Max Planck Institut f\"ur Astronomie, K\"onigstuhl 17, Heidelberg,
  D-69117, Germany} 
\altaffiltext{12}{Department of Physics, Eidgen\"ossische Technische
  Hochschule-Zurich, CH-8093 Zurich, Switzerland}  
\altaffiltext{13}{Department of Physics, Carnegie Mellon University, 5000 Forbes
 Avenue, Pittsburgh, PA 15213} 
\altaffiltext{14}{Institute for Astronomy, University of Edinburgh, Royal
  Observatory, Blackford Hill, Edinburgh EH9 3HJ, U.K. } 
\altaffiltext{15}{Institut d'Astrophysique de Paris, UMR7095 CNRS, Universit\`e 
Pierre et Marie Curie, 98 bis Boulevard Arago, 75014 Paris, France}
\altaffiltext{16}{Observatoire de Paris, LERMA, 61 Avenue de l'Observatoire, 75014
  Paris, France} 
\altaffiltext{17}{Astronomical Institute, Graduate School of Science,
         Tohoku University, Aramaki, Aoba, Sendai 980-8578, Japan}
\altaffiltext{18}{Physics Department, Graduate School of Science, Ehime University,
  2-5 Bunkyou, Matuyama, 790-8577, Japan} 
\altaffiltext{19}{European Southern Observatory, Karl-Schwarzschild-Strasse
  2, D-85748,  Garching, Germany }

\begin{abstract}
We present the results of a search for galaxy clusters in the first 36
XMM-Newton pointings on the COSMOS field. We reach a depth for a total
cluster flux in the 0.5--2 keV band of $3\times10^{-15}$ ergs cm$^{-2}$
s$^{-1}$, having one of the widest XMM-Newton contiguous raster surveys,
covering an area of 2.1 square degrees. Cluster candidates are identified
through a wavelet detection of extended X-ray emission. Verification of the
cluster candidates is done based on a galaxy concentration analysis in
redshift slices of thickness of 0.1--0.2 in redshift, using the multi-band
photometric catalog of the COSMOS field and restricting the search to
$z<1.3$ and i$_{\rm AB}< 25$. We identify 72 clusters and derive their
properties based on the X-ray cluster scaling relations. A statistical
description of the survey in terms of the cumulative $\log(N>S)-\log(S)$
distribution compares well with previous results, although yielding a
somewhat higher number of clusters at similar fluxes. The X-ray luminosity
function of COSMOS clusters matches well the results of nearby surveys,
providing a comparably tight constraint on the faint end slope of
$\alpha=1.93\pm0.04$. For the probed luminosity range of
$8\times10^{42}-2\times10^{44}$ ergs s$^{-1}$, our survey is in agreement
with and adds significantly to the existing data on the cluster luminosity
function at high redshifts and implies no substantial evolution at these
luminosities to $z=1.3$.
\end{abstract}


\section{Introduction}

Clusters of galaxies represent a formidable tool for cosmology (e.g. Borgani
\& Guzzo 2001, Rosati et al. 2002).  As the largest gravitationally relaxed
structures in the Universe, their properties are highly sensitive to the
physics of cosmic structure formation and to the value of fundamental
cosmological parameters, specifically the normalization of the power
spectrum $\sigma_8$ and the density parameter $\Omega_M$.  Clusters are in
principle ``simple'' systems, where the observed properties of the (diffuse)
baryonic component should be easier to connect to the mass of the dark
matter halo, compared to the complexity of the various processes (e.g. star
formation and evolution, stellar and AGN feedback) needed to understand the
galaxy formation.  In particular in the X-ray band, where clusters can be
defined and recognized as single objects, observable quantities like X-ray
luminosity $L_X$ and temperature $T_X$ show fairly tight relations with the
cluster mass (e.g. Evrard et al. 1996, Allen et al. 2001, Reiprich \&
B\"ohringer 2002, Ettori et al. 2004).  Understanding these scaling
relations apparently requires more ingredients than simple heating by
adiabatic compression during the growth of fluctuations (Kaiser 1986; Ponman
et al. 2003, Borgani et al. 2004, 2005). However, their very existence and
relative tightness provides us with a way to measure the mass function
(e.g. B\"ohringer et al 2002, Pierpaoli et al. 2003) and power spectrum
(Schuecker et al. 2003), via respectively the observed X-ray
temperature/luminosity functions and the clustering of clusters, thus
probing directly the cosmological model.  X-ray based cluster surveys, in
addition, can be characterized by a well-defined selection function, which
is an important feature when computing cosmological quantities as first or
second moments of the object distribution.  Finally, and equally important,
clusters at different redshifts provide homogeneous samples of essentially
co-eval galaxies in a high-density environment, enabling studies of the
evolution of stellar populations (e.g. Blakeslee et al. 2003, Lidman et
al. 2004, Mei et al. 2006, Strazzullo et al. 2006).

X-ray surveys in the local ($ \left< z\right> \sim 0.1 $) Universe, stemming
from the ROSAT All-Sky Survey (RASS, Voges et al. 1999) have been able to
pinpoint the cluster number density to high accuracy.  The REFLEX survey, in
particular, has currently yielded the most accurate measurement of the X-ray
luminosity function (XLF, Ebeling et al. 1998; B\"ohringer et al. 2002;
2004), providing a robust $z\simeq 0$ reference frame to which surveys of
distant clusters can be safely compared in search of evolution.
Complementary, serendipitous X-ray searches for high-redshift clusters have
been based mostly on the deeper pointed images from the ROSAT PSPC (RDCS:
Rosati et al. 1995; SHARC: Burke et al. 2003; 160 Square Degrees: Mullis et
al. 2004; WARPS: Perlman et al. 2002) and HRI archives (BMW: Moretti et al.
2004); or on the high-exposure North Ecliptic Pole area of the RASS (NEP:
Gioia et al. 2003; Henry et al. 2006).  A deeper search for massive clusters
in RASS (flux limit $\sim 10^{-12}$ erg cm$^{-2}$ s$^{-1}$) is carried on by
the MACS project (Ebeling et al. 2001), while the XMM-LSS survey is covering
$\sim 9$ square degrees to a flux limit $\sim 10^{-14}$ erg cm$^{-2}$
s$^{-1}$ (Valtchanov et al. 2004)\footnote{All through this paper, we shall
adopt a ``concordance'' cosmological model, with $H_o=70$ km s$^{-1}$
Mpc$^{-1}$, $\Omega_M=0.3$, $\Omega_\Lambda = 0.7$, and --- unless specified
--- quote all X-ray fluxes and luminosities in the [0.5-2] keV band and
provide the confidence intervals on the 68\% level.}. Only recently distant
clusters have started to be identified serendipitously from the Chandra
(Boschin, 2002) and XMM archives, with a record-breaking object recently
identified at $z=1.45$ (Stanford et al. 2006). Overall, these results
consistently show a lack of evolution of the XLF for $L < L^* \simeq 3\cdot
10^{44}$ erg s$^{-1}$ out to $z\simeq 0.8$, At the same time, however, they
confirm the early findings from the Einstein Medium Sensitivity Survey
(EMSS, Gioia et al 1990; Henry et al. 1992) of a mild evolution of the
bright end (Vikhlinin et al. 1998a; Nichol et al. 1999, Borgani et al. 2001;
Gioia et al. 2001; Mullis et al. 2004). In other words, there are
indications that above $z \sim 0.6$ one starts finding a slow decline in the
number of very massive clusters, plausibly indicating that beyond this epoch
they were still to be assembled from the merging of smaller mass units.
These results consistently indicate low values for $\Omega_M$ and
$\sigma_8\sim 0.7$, under very reasonable assumptions on the evolution of
the $L_X-T_X$ relation.  Remarkably, the revised value
$\sigma_8=0.76\pm0.05$ yielded by the recent 3-year WMAP data (Spergel et
al. 2006), is in close agreement to those consistently indicated by all
X-ray cluster surveys over the last five years, both from the evolution of
the cluster abundance (Borgani et al. 2001; Pierpaoli et al. 2003; Henry
2004) and the combination of local abundance and clustering (Schuecker et
al. 2003).

The precision achievable in these calculations essentially reflects the
uncertainty in the relation between cluster mass and measurables such as
$L_x$ or $T_x$.  Progress in the knowledge of these relations and their
evolution is currently limited by the small number of clusters known at high
redshifts (less than 10 objects known at $z>1$), although important results
have been recently achieved (Ettori et al. 2004; Maughan et al. 2006; Kotov
\& Vikhlinin 2005). In addition, virtually all current statistical samples
of distant clusters have been selected serendipitously from sparse archival
X-ray images, and reach maximum depths of $\sim 10^{-14}$ erg cm$^{-2}$
s$^{-1}$.  The former limits the ability to study distant clusters in the
context of their surrounding large-scale structure; the latter limits to
relatively low redshifts the study of low-luminosity groups.

The Cosmic Evolution Survey (COSMOS), covering 2 \sq\deg, is the first HST
survey specifically designed to thoroughly probe the evolution of galaxies,
AGN and dark matter in the context of their cosmic environment (large scale
structure -- LSS). COSMOS provides a good sampling of LSS, covering all
relevant scales out to $\sim$ 50 -- 100 Mpc at z $> 0.5$ (Scoville et
al. 2007a). The rectangle bounding all the ACS imaging has lower left and
upper right corners (RA,DEC J2000) at (150.7988\deg,1.5676\deg) and
(149.4305\deg, 2.8937\deg). To define the LSS, both deep photometric
multi-band studies of galaxies are carried out (Scoville et al. 2007b) as
well as an extensive spectroscopy program (Lilly et al. 2007).

The 1.4 Msec XMM-Newton observations of the COSMOS field (Hasinger et
al. 2007) provide coverage of an area of 2.1 \sq\deg\ to unprecedented depth
of $10^{-15}$ ergs cm$^{-2}$ s$^{-1}$, previously reached by deep surveys
only over a much smaller area (Rosati et al. 2002).
At the same time, the large multi-wavelength coverage makes optical
identification very efficient (Brusa et al. 2007; Trump et al. 2007). The
zCOSMOS program (Lilly et al. 2007) and the targeted follow-up using the
Magellan telescope (Impey et al. 2007) are of particular importance. While
we anticipate a completion of these programs within the next years, we
present in this paper the cluster identification based on the photometric
redshift estimates, available as a result of an extensive broad-band imaging
of the COSMOS field (Capak et al. 2007; Taniguchi et al. 2007; Mobasher et
al. 2007).

By design, this paper concentrates on the densest parts of the LSS, which
have already collapsed to form virialized dark matter halos of groups and
clusters of galaxies, populated with evolved galaxies. This approach is
quite complementary to the study of extended supercluster-size structures,
which is presented in Scoville et al. (2007b).

\begin{figure*}
\includegraphics[width=16.cm]{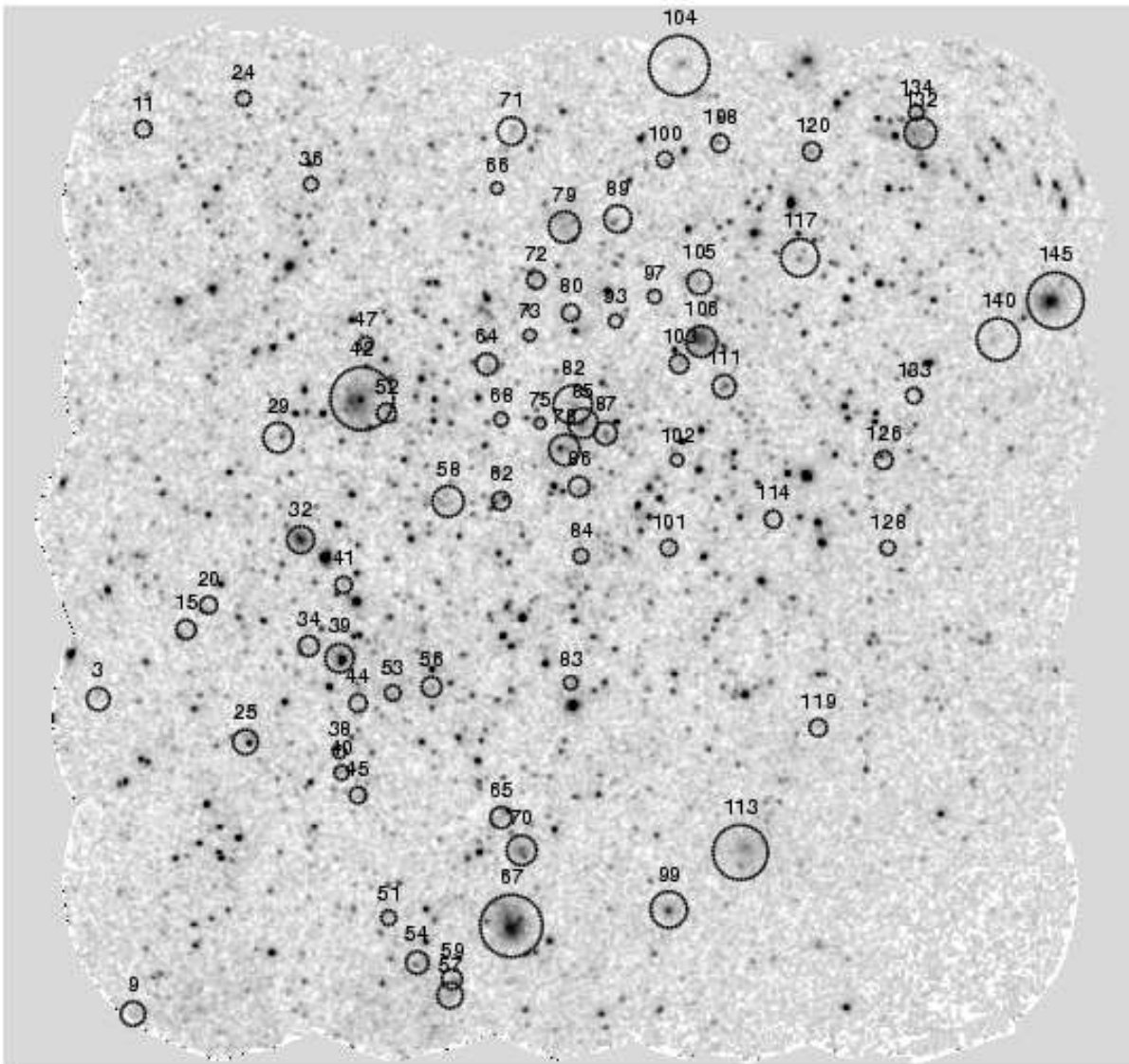}

\figcaption{An image of the signal-to-noise ratio in the 0.5--2 keV band,
smoothed with a Gaussian of two pixel width. White color corresponds to
values smaller than -0.5, while black corresponds to values exceeding
3. Circles indicate the position and size of the detected clusters,
labeled in correspondence with the cluster number in the catalog. The image
is 1.5 degrees on a side. The pixel size is $4^{\prime\prime}$ on a side.
A tangential elongation in the point spread function of XMM-Newton
telescopes is clearly visible in the NW corner of the image, which has the
largest exposure time.
\label{f:s2n}}

\end{figure*}

The paper is organized as follows: \S\ref{data} describes the analysis of
the XMM-Newton observations of the COSMOS survey; \S\ref{search} presents
our diffuse source detection technique; \S\ref{sec-cat} describes the
wavelet analysis of the photometric galaxy catalog; \S\ref{xcat} provides
the catalog of the identified X-ray groups and clusters of galaxies;
\S\ref{xfun} derives the luminosity function; and \S\ref{resume}
concludes the paper.

\section{XMM-COSMOS Data reduction}\label{data}

For cluster detection, we used the XMM-Newton mosaic image in the 0.5--2 keV
band, based on the first 36 pointings of the XMM-Newton observations of the
COSMOS field. The first 25 pointings completely cover the 2 square degree
area of COSMOS at 57\% of its final depth (Hasinger et al. 2007). 11
pointings have already been obtained as a part of the second year
observations and are included in the present analysis. A description of the
XMM-Newton observatory is given by Jansen et al. (2001). In this paper we
use the data collected by the European Photon Imaging Cameras (EPIC): the
{\it pn}-CCD camera (Str\"uder et al. 2001) and the MOS-CCD cameras (Turner
et al. 2001).  All Epic-{\it pn} observations have been performed using the
Thin filter, while both Epic-MOS cameras used the Medium filter. An analysis
of absolute frame registration for the XMM COSMOS survey has been carried
out by Cappelluti et al. (2007). The offsets do not exceed
$3^{\prime\prime}$ and do not affect the current analysis.

In addition to the standard data processing of the EPIC data, which was done
using XMMSAS version 6.5 (Watson et al. 2001; Kirsch et al. 2004; Saxton et
al. 2005), we perform a more conservative removal of time intervals affected
by solar flares, following the procedure described in Zhang et
al. (2004). In order to increase our capability of detecting extended, low
surface brightness features, we have developed a sophisticated background
removal technique, which we coin as 'quadruple background subtraction',
referring to the four following steps:

First, we remove from the image accumulation the photons in the energy band
corresponding to the Al $K_{\alpha}$ line for {\it pn} and both MOS
detectors and the Si $K_{\alpha}$ line for both MOS detectors.  The
resulting countrate-to-flux conversion in the 0.5--2 keV band excluding the
lines is $1.59\times10^{-12}$ for {\it pn} and $5.41\times10^{-12}$ for each
MOS detector, calculated for the source spectrum, corresponding to the APEC
(Smith et al. 2001) model for a collisional plasma of 2 keV temperature, 1/3
solar abundance and a redshift of 0.2.

Second, we subtract the out-of-time events (OOTE) for {\it pn}. In order to
do so, we generate an additional event file for each {\it pn} event file,
which entirely consists of events emulating the OOTE, produced using the
XMMSAS task {\it epchain}. We then apply the same selection of events using
the good time intervals, generated during the light curve cleaning of the
main event file. The last step is to extract the images with the same
selection criteria (energy, flag, event pattern), normalize them by the
fraction of OOTE (0.0629191 for the Full Frame readout mode used) and to
subtract them from the main image.

Third and fourth, we use two known templates for the instrumental
(unvignetted) and the sky (vignetted) background (Lumb et al. 2002,
2003). The instrumental background is caused by the energetic particles and
has a uniform distribution over the detector. The sky background consists of
the foreground galactic emission as well as unresolved X-ray background and
its angular distribution on scales of individual observation could be
considered as flat on the sky, therefore following the sensitivity map of
the instrument (Lumb et al. 2002). To calculate the normalization for each
template, we first perform a wavelet reconstruction (Vikhlinin et al. 1998b)
of the image without a sophisticated background subtraction. We excise the
area of the detector, where we detect flux on any wavelet scale and then
split the residual area into two equal detector parts with higher and lower
values of vignetting than the median value. By imposing a criterion that the
weighted sum of the two templates should reproduce the counts in both areas,
we obtain a system of two linear equations for two weighting
coefficients. By solving the latter we derive the best estimate for the
normalizations of both templates.

After the background has been estimated for each observation and each
instrument separately, we produce the final mosaic of cleaned images and
correct it for the mosaic of the exposure maps in which we account for
differences in sensitivity between pn \& MOS detectors stated above.  The
detailed treatment of the background is newly developed and has been adopted
for all XMM-COSMOS projects (Cappelluti et al. 2007). The procedure of
mosaicing of XMM-Newton data has been used previously and is described in
detail in Briel et al. (2004).

The resulting signal-to-noise ratio image is shown in Fig.~\ref{f:s2n}. As
can be seen from the figure, the image exhibits a fairly uniform
signal-to-noise ratio. Without the refined background subtraction, the
signal-to-noise image exhibited large-scale variations, which could mimic an
extended source. On the image, the circles show the position and the angular
extent of detected clusters of galaxies. The pixel size of the X-ray images
employed in the current analysis is $4^{\prime\prime}$ on a side.

\begin{figure*}
\includegraphics[width=17.cm]{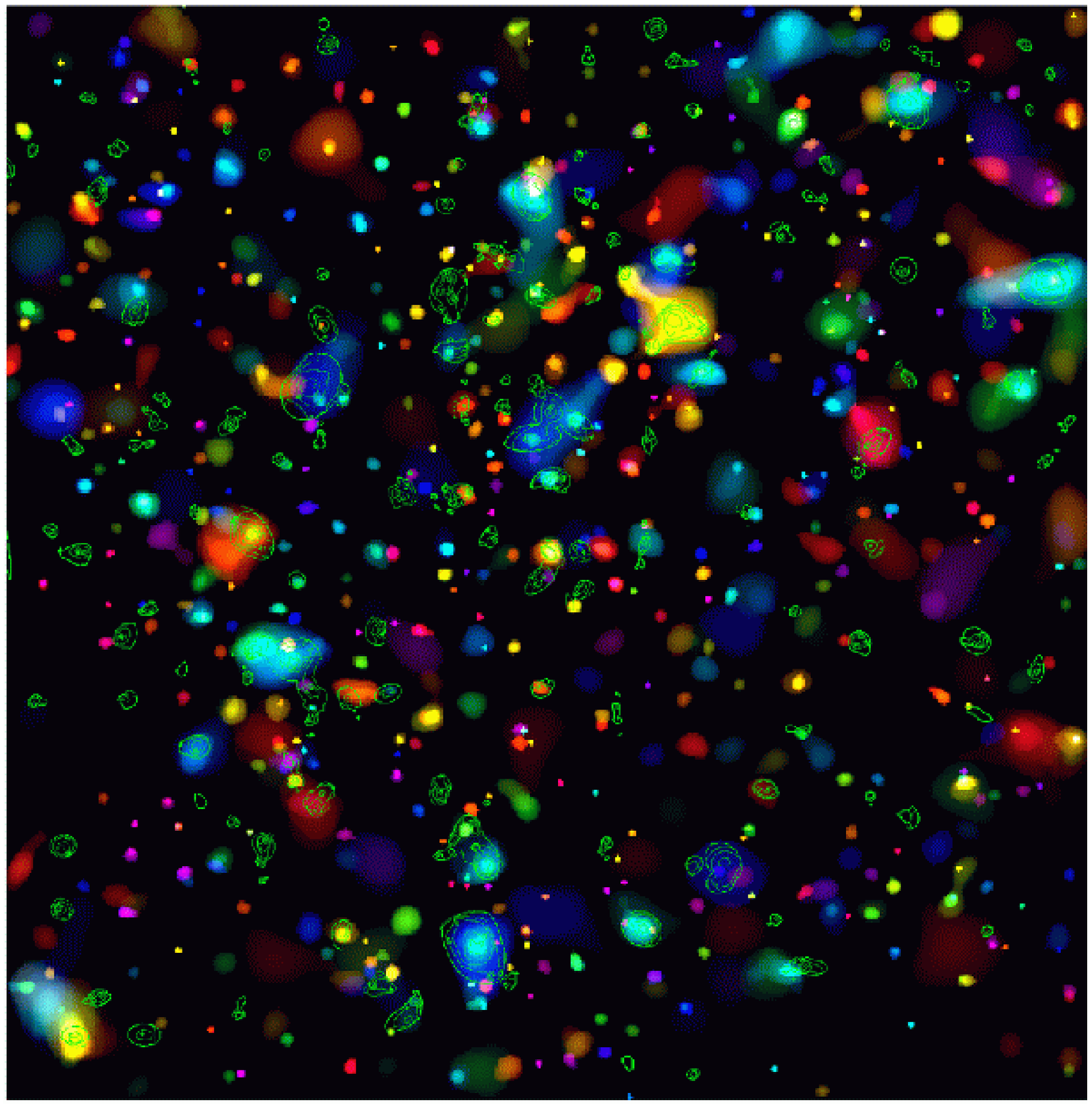}
\figcaption{
  The colors of COSMOS. The wavelet reconstruction of the early-type galaxy
  concentrations searched in the photo-z catalog is color-coded according to
  the average redshift: blue -- 0.2, cyan -- 0.4, green -- 0.6, yellow --
  0.8, red -- 1.0. The green contours outline the area of the X-ray emission
  associated with 150 extended source candidates. The image is 1.5 degrees
  on a side. The pixel size is $10^{\prime\prime}$ on a side.
\label{f:coscol}}
\end{figure*}

\section{Cluster detection technique}\label{search}

We used the wavelet scale-wise reconstruction of the image, described in
Vikhlinin et al. (1998b), employing angular scales from $8^{\prime\prime}$
to $2.1^\prime$. We apply the approach of Rosati et al. (1998), Vikhlinin et
al. (1998b), and Moretti et al. (2004) to effectively select clusters of
galaxies by the spatial extent of their X-ray emission. The cluster
detection algorithm consists of two parts: (1) selection of the area with
detectable flux on large angular scales, and (2) removal of the area where
the flux could be explained by contamination from embedded point-like
sources, most of which are background AGNs, according to their optical
identification (Brusa et al. 2007). When applying this approach to
XMM-Newton data, the off-axis behavior of the point-spread function (PSF)
(e.g.  Valtchanov et al. 2001) needs to be taken into account. The major
change in the two-dimensional shape of the PSF of the XMM telescope is a
tangential elongation with respect to the optical axis of the telescope
(Lumb et al. 2003). In the wavelet scale-wise reconstruction of the point
source, this produces a flux redistribution between the $8^{\prime\prime}$
and $16^{\prime\prime}$ scales with more flux going into the
$16^{\prime\prime}$ at large off-axis angles. At the same time, the flux
contained in the sum of $8^{\prime\prime}$ and $16^{\prime\prime}$ wavelet
scales does not vary much with the off-axis angle and its ratio to the flux
detected on larger angular scales is greater than 1.9 at any off-axis angle.
Smaller flux ratios have therefore been chosen as a detection criterion for
an extended source.

Our wavelet algorithm (Vikhlinin et al. 1998b) generates a noise map against
which the flux in the image is tested for significance. We assess the
sensitivity of the survey, by examining the noise map corresponding to the
$32^{\prime\prime}$ scale. The total area over which the source of a given
flux can be detected is the area where the source signal to noise ratio
exceeds the chosen detection threshold of 4. For each cluster this number is
unique, yet the total flux of the cluster is not recovered at
$32^{\prime\prime}$. Any procedure to recover the total flux introduces a
scatter in the total flux vs survey area relation. The noise map is
calculated using the data, and so it includes the effect of decreasing
sensitivity to a diffuse source detection caused by the contamination from
other X-ray sources. Our calculation of the sensitivity neglects
cluster-cluster and cluster-point source correlations. In this regard, a
presence of an unresolved bright cool core could complicate the detection of
a cluster. A modelling of such effect will be done within the granted
C-COSMOS program.

The source appearance on the $32^{\prime\prime}$ wavelet scale defines the
area which is further used for measuring the parameters of the source. The
total flux of the source is obtained by extrapolation of the flux measured
within this area. As a first step in selecting the cluster candidates, we
clean our source list from contamination due to point sources. For that, we
compare the sum of the two smallest scales ($8^{\prime\prime}$ and
$16^{\prime\prime}$) with the sum of the largest scales
($32^{\prime\prime}$, $64^{\prime\prime}$, $128^{\prime\prime}$) in the same
area to decide whether the area {\it contains} a diffuse source. If the sum
of the two smallest scales exceeds the sum of the three largest scales by a
factor of 1.9 or greater, the source is removed from the diffuse source
list. This procedure allows us to deal with any number of point sources
embedded in the same area. We find that 80\% of the initially selected areas
are dominated by the flux from point sources. Out of the remaining 150
diffuse source candidates, surviving this test, we identify only 76 with
peaks in the galaxy distribution, as detailed in \S\ref{sec-cat}. The bulk
of unidentified sources are confused low-luminosity AGNs, which are not
detected on the small scales. Some of those sources might be identified by
chance with a galaxy overdensity. The probability of such event is 1\%, as
determined by the fraction of the total XMM-COSMOS area occupied by the
unidentified sources, which were initially considered as extended source
candidates. Sect.~\ref{sec-cat} describes the construction of the catalog of
420 galaxy groups based on the multiband photometric data.  The number of
optical peaks missing identification with an extended X-ray source is 345,
leading to a chance identification for $\sim3$ clusters. The
misidentifications have large separations between the X-ray center and the
optical center of a group, as a chance coincidence of the centers is
negligibly small. So, through a comparison to optical images we removed 4
sources whose shape of X-ray emission was not associated with a group of
galaxies, finally reducing the cluster sample size to 72. For all four
rejected systems, one can decompose the emission into 3-5 unresolved AGN
using the K-band images identification, similar to the primary method for
AGN identification used in the COSMOS field (Brusa et al.  2007).

\begin{figure*}
\includegraphics[width=17.cm]{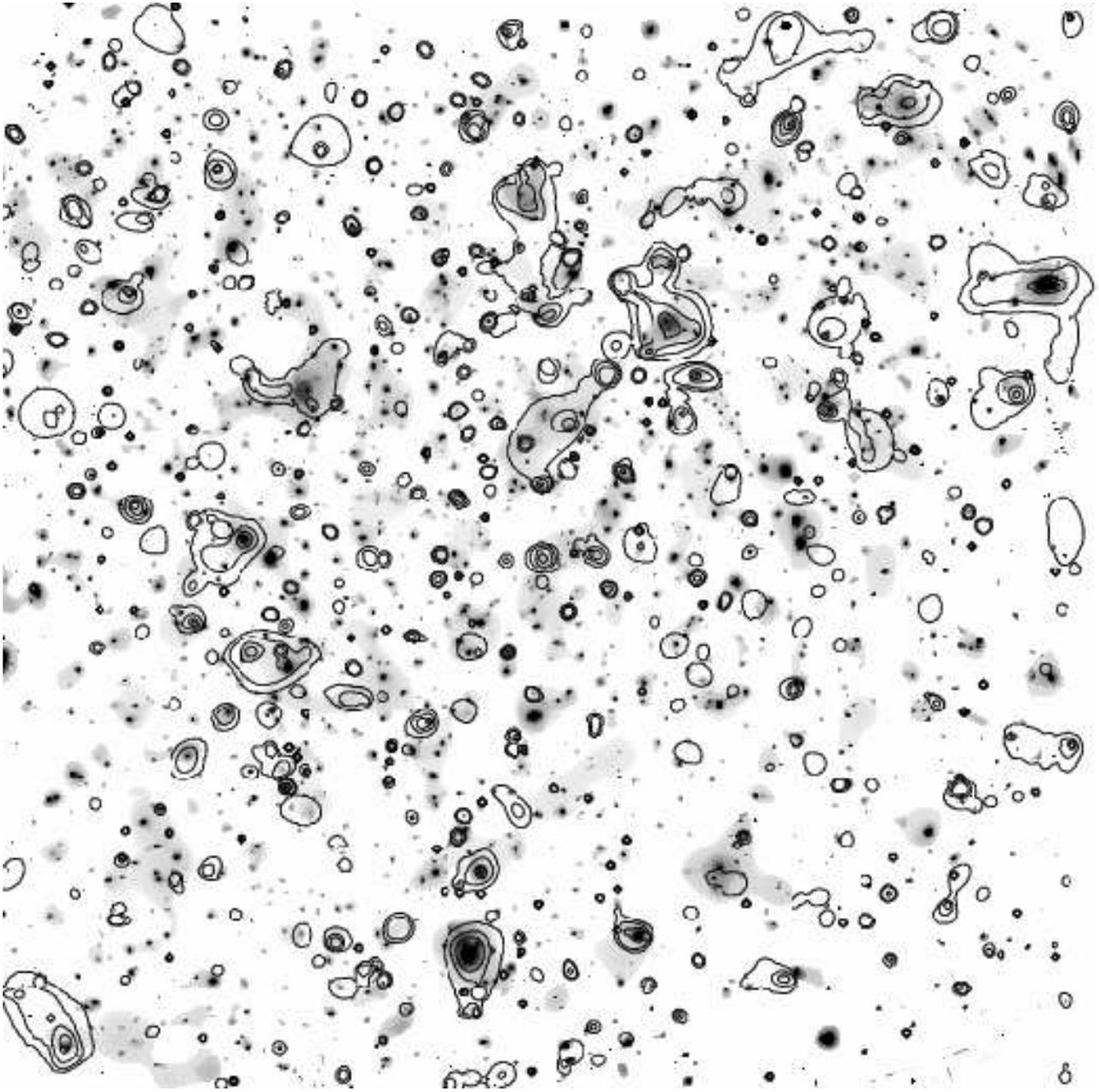} 
\figcaption{
Scale-wise wavelet reconstruction of the XMM-COSMOS mosaic image in the
0.5--2 keV band. Grey color corresponds to the surface brightness of
$10^{-7}$ counts s$^{-1}$ pixel$^{-1}$. A surface brightness level exceeding
$10^{-5}$ counts s$^{-1}$ pixel$^{-1}$ is shown in black. Contours indicate
the location and strength of galaxy structures, identified in the photo-z
catalog. The image is 1.5 degrees on a side. The pixel size is
$4^{\prime\prime}$ on a side.
\label{f:x2ph}}

\end{figure*}

It is possible to reduce the number of spurious X-ray detections. However,
it would result in a loss of survey sensitivity (Vikhlinin et al. 1998b) and
is usually done to improve the efficiency of the optical follow-up. Since
such a follow-up already exists in the COSMOS field, we prefer to keep the
high sensitivity achieved by our detection method.

Although some additional flux is detected on larger scales, the source
confusion becomes large on scales exceeding $1^{\prime}$ at the flux limits
typical of our survey. In addition, the use of only a single scale for the
detection has the advantage of allowing a more straight-forward modeling.

Once a diffuse source is detected, the next step is to estimate its total
flux. Unlike in the case of point sources, this is complicated by the
unknown shape of a diffuse source. For bright sources, it is possible to
carry out a surface brightness analysis and estimate the missing flux
directly. For example, an analysis of the surface brightness profile for a
bright cluster at $z=0.22$ in COSMOS using the beta model (Smol{\v c}i{\' c}
et al. 2007) results in a factor of 1.8 higher flux compared to the total
flux within the aperture defined by the source detection procedure.

Studies of nearby clusters of galaxies (e.g. Markevitch 1998) show that most
of the cluster flux is encompassed within the radius $r_{500}$ (a radius
encompassing the matter density 500 times the critical), so we adopt this
radius for the total flux measurements, $F(<r_{500})$. We have therefore
adopted a method for estimating $r_{500}$, based on the iteratively
corrected observed flux

\begin{equation}
F(<r_{500})=C_{\beta}(z,T) F_{\rm d}
\end{equation}

with the correction, $C_{\beta}(z,T)$, defined iteratively through the
scaling relations as described in \S\ref{flux} and using the flux and
redshift of the system. To estimate the observed flux we use the total
counts ($F_{\rm d}$) in the area defined by the detection algorithm ($R_{\rm
t}$) and subtract the contribution from the embedded point sources based on
intensity determined by the $4^{\prime\prime}$ scale
($R_{4^{\prime\prime}}$) of the wavelet. The total flux retained in the area
is larger than $R_{4^{\prime\prime}}$ by a factor, determined by the size of
the area ($a$) and the XMM PSF ($C_{\rm PSF}(a)$), with values in the range
1.1--1.5. We work in units of MOS1 counts/s on the mosaic. We add pn and MOS
counts as observed, while putting more effective weight to the pn exposure
map, so that
\begin{equation}
F_{\rm d}= 5.41
\times 10^{-12} (R_{\rm t}-C_{\rm PSF}(a)\times R_{4^{\prime\prime}})
{\rm ergs\; s}^{-1} {\rm cm}^{-2}
\end{equation}
Also, after removal of cool cores, the scatter in the scaling relations is
found to be moderate (Markevitch et al. 1998; Finoguenov et al. 2005a). In
our survey, a cool core is indistinguishable from an AGN and is therefore
removed from the flux estimates.
%

As has been described above, the area for cluster flux extraction has been
selected based on a particular ($32^{\prime\prime}$) spatial scale. Thus,
only a part of the cluster emission is used in the detection. This results
in an effective loss of sensitivity, as not all the cluster emission is
used. However, the modeling of the detection becomes simple and in
particular, an influence of the cool cluster cores on the cluster detection
statistics is reduced.

\section{Use of photometric redshifts to identify the clusters}\label{sec-cat}

The identification/confirmation of a bound galaxy system would be most
robust through optical/NIR spectroscopic redshift measurement, but this is
costly for such a large survey. Photometric redshifts, when measured with
sufficient accuracy, represent a viable alternative. The COSMOS survey,
providing both the photo-z (Mobasher et al. 2007) and in the future the
spectral-z (Lilly et al. 2007) is an ideal field to understand all the pros
and cons of the use of photo-z to identify the optical counterparts to X-ray
clusters. Diffuse X-ray emission, associated with a galaxy overdensity, is
by itself a proof of the virialized nature of the object (Ostriker et
al. 1995), so a combination of diffuse X-ray source and a photo-z galaxy
overdensity validates the photo-z source as a virialized object. Deep
multiband photometric observations have been carried out in the COSMOS field
using CFHT, Subaru, CTIO and KPNO facilities (Capak et al. 2007). The
combination of depth and bands allows us to produce an $i$-band based
photo-z catalog with a $1\sigma$ uncertainty of the redshift estimate of
$0.027 (1+z)$ for galaxies with $i_{AB}< 25$ (Mobasher et al. 2007) obtained
without the luminosity prior. While both z and K-band data are used in
determining the photometric redshifts, the input catalog is based on the
$i$-band images. Therefore, the redshift range of the cluster search in this
paper has to be limited to redshifts below 1.3, after which the 4000\AA\
break moves redward of the $i$-band filter.

To provide an identification of galaxy overdensities in the photo-z space,
we select from the original photo-z catalog only the galaxies, that (1) are
classified in their SED (spectral energy distribution) as early-types
(ellipticals, lenticulars and bulge-dominated spirals); (2) have high
quality photo-z redshift determination (95\% confidence interval for the
redshift determination is lower than $\Delta z=0.4$); (3) are not
morphologically classified as stellar objects. Given the current quality of
photo-z, we select redshift slices covering the range $0<z<1.3$ with both
$\Delta z=0.1$ and $\Delta z=0.2$. To provide a more refined redshift
estimate for the identified structures, we slide the selection window with a
0.05 step in redshift. We add each galaxy as one count and apply wavelet
filtering (Vikhlinin et al. 1998b) to detect enhancements in the galaxy
number density on scales ranging from 20$^{\prime\prime}$ to 3$^{\prime}$ on
a confidence level of $4\sigma$ with respect to the local background.  The
local background itself is determined by both the field galaxies located in
the same redshift slice as well as galaxies attributed to the slice due to a
catastrophic failure in the photo-z. The fraction of galaxies assigned to
structures in our analysis varies from 15\% at a $z=0.3$ to 2\% at $z=1.2$.
However, the level of the background does not vary strongly with redshift,
so the detection of the optical counterpart is primarily determined by the
sensitivity, required to determine the photometric redshift for sufficient
number of members. As will be discussed below, the X-ray detection limits at
$z>1$ in the COSMOS survey correspond to a massive cluster, that typically
has a sufficient number of bright members, enabling the success of the
identification procedure.

The angular scales selected for the analysis match the extent of X-ray
sources found. While the number of galaxies within each cluster is not
determined by the statistics, our ability to see it over the background is
statistical with the errors determined by the background level. Analysis of
galaxy overdensities is not sensitive to catastrophic failures in the
photo-z catalog, as those simply reduce the strength of galaxy
concentration. Larger structures, selected without the prior on the SED
type, are reported in Scoville et al. (2007b) and should be understood as
the LSS environment within which the high-density peaks detected via X-ray
emission are located. Making a prior on SED type is technically necessary to
increase the contrast in the photometric data. From a theoretical point of
view, the use of these specific SEDs is justified by the large ages of
cluster ellipticals and the lack of evolution in their morphological
fraction with redshift (Smith et al. 2005; Postman et al. 2005; Wechsler et
al. 2005; see also direct estimates from the COSMOS field in Capak et
al. 2007 and Guzzo et al. 2007), where the already tested redshift range
reaches $z=1.1$ (Hashimoto et al. 2005). The ultra-deep field of the UKIRT
Infrared Deep Sky Survey (Lawrence et al. 2006) finds that the galaxy red
sequence disappears beyond $z=1.5$ (Cirasuolo et al. 2006).  Moreover, local
spiral-rich groups also do not reveal X-ray emission associated with their
IGM (Mulchaey et al. 1996). In \S\ref{xfun} we verify the completeness of
the source identification using the $V/V_{max}$ test and conclude that we do
not lack high-redshift identification in our analysis. In comparison with
the widely adopted cluster red sequence method, the use of photo-z data
allows us to include fainter members, which is of particular importance for
galaxy groups.

\include{cat_tab4}

In Fig.~\ref{f:coscol} we overlay the X-ray contours of diffuse sources over
the color-coded image of the structures identified via the photo-z galaxy
selection. The brightness of the color is proportional to the number density
of galaxies, while the color represents the average redshift. In addition,
to show the correspondence between galaxy structures and X-ray emission, we
provide in Fig.~\ref{f:x2ph} an overlay of the wavelet-reconstructed X-ray
image with the contours, showing the sum of the wavelet-filtered redshift
galaxy slices.

To determine the redshift of an X-ray structure, we calculate the position
and maximum of the galaxy density peak using the photo-z slices. The
extended source candidate is considered identified as a cluster if it
contains a galaxy density peak inside the detected extend, which is
typically $20^{\prime\prime}-40^{\prime\prime}$. The redshift and center of
the cluster is selected to be associated with the strongest galaxy peak
inside the cluster. Due to the wings in the redshift estimate using the
photometric technique, the cluster is detected over a range of
redshifts. Selecting the peak value is thought to yield the most likely
redshift of the cluster. In the case of several overlapping structures, this
could lead to biases due to redshift selection effects and in any case
should be verified. In one instance (cluster 133), we manually put a
preference for a high redshift counterpart, which matched both the X-ray
center and the redshift of BCG of the system. In total, 72 X-ray clusters
have been identified using this method. Their properties are listed in
Tab.~1 and discussed below.

\section{A catalog of identified X-ray clusters}\label{xcat}

In this section we describe our catalog of 72 X-ray galaxy clusters detected
so far in the COSMOS field. In the catalog we provide the cluster
identification number (column 1), R.A. and Decl. of the peak of the galaxy
concentration identified with the extended X-ray source in Equinox J2000.0
(2--3), the estimated radius $r_{500}$ in arcminutes (4), the cluster flux
in the 0.5--2 keV band within $r_{500}$ in units of $10^{-14}$ ergs
cm$^{-2}$ s$^{-1}$ with the corresponding 1 sigma errors (5), photometric
redshift (6), an estimate of the cluster mass, $M_{500}$, in units of
$10^{13} M_\odot$ (7), rest-frame luminosity in the 0.1--2.4 keV band (8),
an estimate of the IGM temperature in keV (9), the source of redshift (10)
with 1 denoting the photometric redshift estimates and 0 denoting the
redshift refinement using information available from the SDSS DR4 (York et
al.  2000, Abazajian et al. 2005; Adelman-McCarthy et al. 2006), the results
of the pilot zCOSMOS program, and the first results of the main zCOSMOS
program (Lilly et al. 2007). The refinement of the photo-z is complete for
$z<0.2$, which reduces the degree of uncertainty in calculation of
distance-dependent properties of the sample to a level, well within the
scatter in the scaling relations for X-ray clusters (e.g. Stanek et
al. 2005). The formal error on the positional uncertainty is of the order of
$10^{\prime\prime}$. The errors provided on the derived properties are only
statistical. A future study of systematic errors will be done by direct
comparison between our reported mass estimates and the weak lensing
measurements (Leauthaud et al., in prep.; Taylor et al., in prep.). For a
number of clusters with best X-ray statistics, it appears that the peak of
the galaxy density can be somewhat offset from the X-ray center, which in
turn has a better match with the position of the brightest cluster
galaxy. Further investigation of this interesting property will be done
elsewhere.

In order to calculate the properties of the clusters, we use the known
scaling relations at low-redshift and evolve them according to the recent
Chandra and XMM observations of high-redshift clusters of galaxies (Kotov \&
Vikhlinin 2005; Finoguenov et al. 2005b; Ettori et al. 2004; Maughan et
al. 2006). Following Evrard et al. (2002), we choose not to correct for the
redshift evolution of the ratio of the density encompassed by the cluster
virial radius to the critical density.

\begin{figure*}
\includegraphics[width=7.5cm]{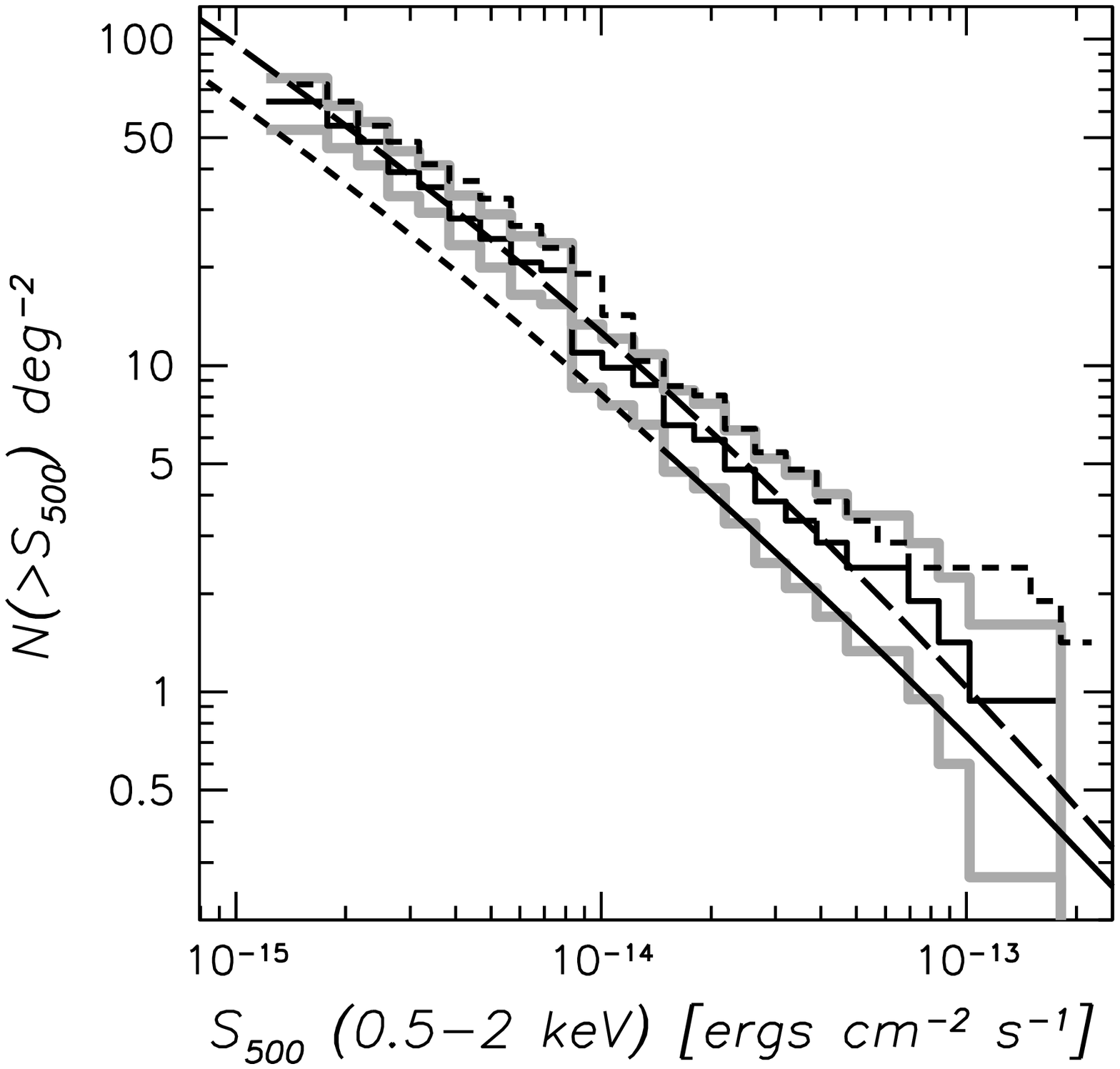}\hfill
\includegraphics[width=7.5cm]{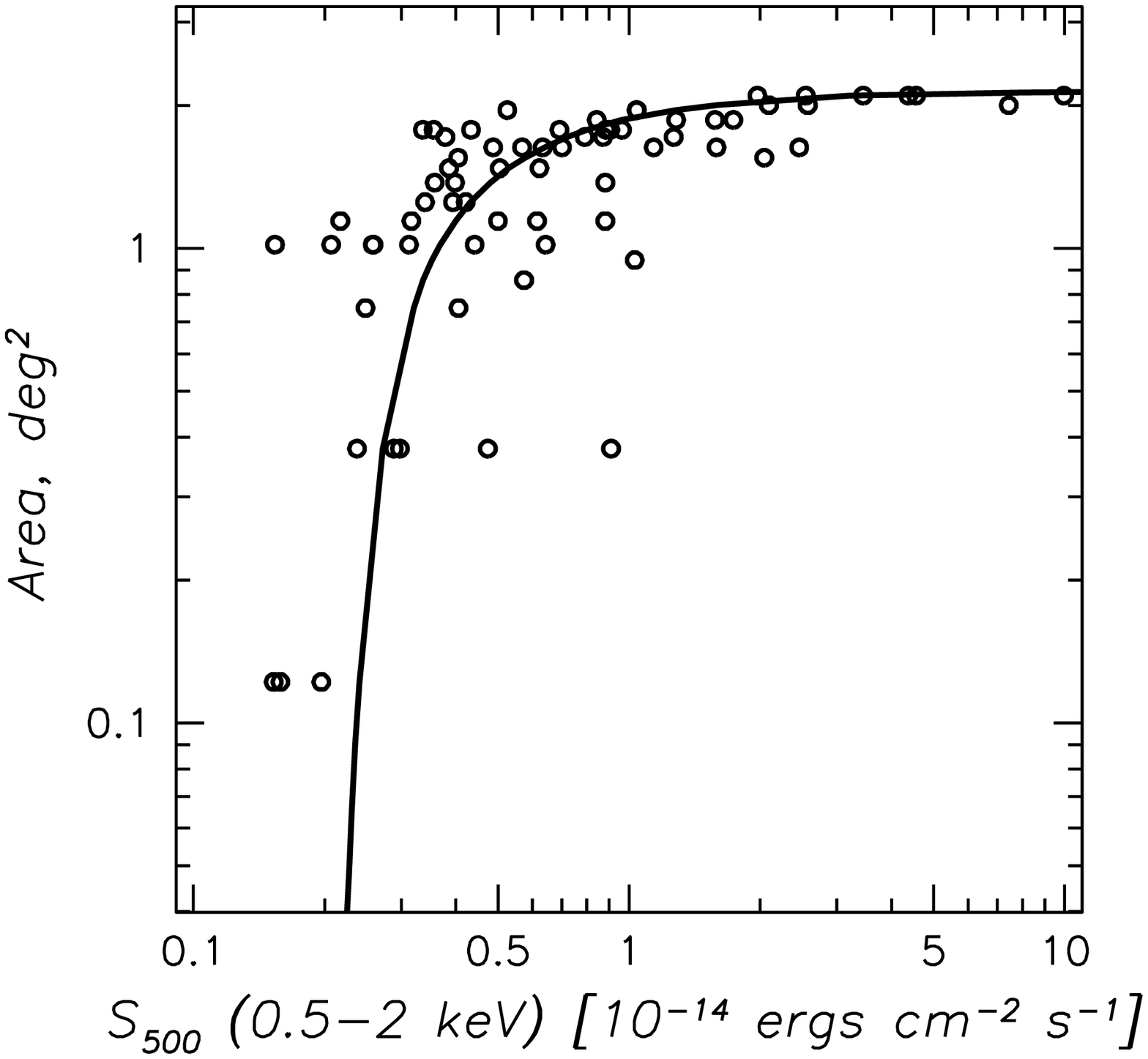}

\figcaption{{\it Left panel}: Cumulative cluster number counts
($\log(N>S)-\log(S)$) for the COSMOS field. The solid histogram shows the
data and grey histograms denote the 68\% confidence interval. The black
solid/short-dashed curve shows the results of the modeling of RDCS (Rosati
et al. 2002), with the solid part corresponding to fluxes sampled by their
data, while the short dashed line denotes the model prediction. The long
dashed curve shows the prediction for no evolution in the luminosity
function in Rosati et al. (2002), which provides a good fit to our data. The
dashed histogram shows a typical difference due to assumption of the scaling
relations. {\it Right panel}: Survey area as a function of the total source
flux in the 0.5--2 keV band. Open circles show the area corresponding to the
source, while the solid line shows the sensitivity curve for the flux in the
detection cell, scaled up to match the displayed data. The scatter of points
around the curve is determined by differences in the flux losses between
distant and nearby sources.
\label{f:logn}}

\end{figure*}

\subsection{Flux estimates}\label{flux}

Even a basic statistical description of the cluster sample, such as
$\log(N)-\log(S)$, is not entirely straight-forward, as it requires a
knowledge of the level of the emission at large distances from the cluster
center. Most cluster surveys to date either use much deeper reobservation of
the clusters for this purpose or extrapolate the detected cluster flux
assuming some model for the spatial shape of the emission. Although most
observers agree that the $\beta$-model (Jones \& Forman 1984; 1999) is a
good description of the cluster shape, characterization of the outskirts of
groups and clusters of galaxies is still uncertain (e.g. Vikhlinin et
al. 2006; Borgani et al. 2004). Moreover, groups reveal a large scatter in
the shape of their emission (Finoguenov et al. 2001; Mahdavi et al. 2005).

For the purpose of this paper, we extrapolate the measured flux ($F_{\rm
d}$) by an amount of $C_{\beta}(z,T)$, which uses the $\beta$-model
characterization of cluster emission, with the following parameters
\begin{equation}
\beta=0.4\times ({kT / 1 {\rm keV}})^{1/3}
\end{equation}
and 

\begin{equation}
r_{\rm core}=0.07\times ({kT / 1 {\rm keV}})^{0.63}\times r_{500}
\end{equation}

We note that although the cool core clusters require a second
spatial component to describe the central peak of their emission, in our
case the contribution of a cool core cannot be distinguished from point
sources and has therefore been removed from the flux estimates. Since only
the large-scale component exhibits a scaling with temperature (Markevitch
1998), removal of the center should result in reduced scatter in the derived
cluster characteristics, such as the total mass. More importantly, exclusion
of cool cores from the detection procedure decreases an observational bias
towards low-mass groups with strong cool cores, and make a selection of the
sample to be closer to mass selection, and reduces the bias discussed in
Stanek et al. (2006) and O'Hara et al. (2006), while facilitating the
modelling of the cluster detection. However, subtle differences in the
derived characteristics of the sample are expected as discussed below. In
calculating the rest-frame luminosity, we iteratively use the total flux
within an estimated $r_{500}$ ($C_{\beta}(z,T) F_{\rm d}$) and apply the
K-correction ($K(z,T)$) accounting for the temperature and redshift of the
source, following the approach described in B\"ohringer et al. (2004) and
assuming an element abundance of 1/3 solar. So, finally, the derived
luminosity is
\begin{equation}
L_{0.1-2.4 \rm keV} = 4\pi d_L^2 K(z,T) C_{\beta}(z,T) F_{\rm d}
\end{equation}

We note that a similar approach for reconstructing the cluster flux is taken
in Henry et al. (1992), Nichol et al. (1997), Scharf et al. (1997), Ebeling
et al. (1998), Vikhlinin et al. (1998b), de Grandi et al. (1999), Reprich \&
B\"ohringer (2002), and B\"ohringer et al. (2004). A choice of $r_{500}$ as
a limiting radius is motivated by the results of Vikhlinin et al. (2006) on
the steepening in the surface brightness profiles, observed near this
radius.

To estimate the temperature of each cluster, we use the $L_{0.1-2.4 \rm
keV}-T$ relations of Markevitch (1998) for the case of excised cool cores:
\begin{equation}
kT=6\; {\rm keV} \times 10^{(\log_{10}(L_{0.1-2.4 \rm keV}
E_z^{-1})-44.45)/2.1)}
\end{equation}
where 
\begin{equation}
E_z=(\Omega_M (1+z)^3 +
\Omega_\Lambda)^{1/2}
\end{equation}

The estimates of the total gravitational mass and the corresponding
$r_{500}$ are performed using the M--T relation (F. Pacaud 2005, private
communication) re-derived from Finoguenov et al. (2001) using an orthogonal
regression and correcting the masses to $h_{70}$ and a $\Lambda$CDM
cosmology: 
\begin{equation}
M_{500}=2.36\times10^{13}M_\odot \times T^{1.89}
E_z^{-1}
\end{equation}
 and 

\begin{equation}
r_{500}=0.391\, {\rm Mpc} \times (kT/{\rm keV})^{0.63}E_z^{-1}
\end{equation}

Although the $L_{500}-T$ relation of Markevitch (1998) is formally derived
for high-temperature clusters, a comparison of that relation with the new
results on the $L_{500}-T$ relation for groups (Ponman T.J., private
communication), shows that the two are very similar.  Based on direct
comparison between the predicted and measured temperatures for a number of
clusters in the COSMOS (e.g. Smol{\v c}i{\' c} et al. 2007; Guzzo et
al. 2007), the mass estimates provided in Tab.~1 should be good to a factor
of 1.4, due to both the scatter in the scaling relations and uncertainty in
our knowledge of their redshift evolution.

\subsection{Cluster counts}

It is common to characterize a cluster survey by its area as a function of
the limiting flux and present the results as a relation between a cumulative
surface density of clusters above a given flux limit vs the flux value, the
cluster $\log(N>S)-\log(S)$ (e.g. Rosati et al. 1998). In order to take into
account the difference between the total flux and the observed flux, the
$\log(N>S)-\log(S)$ is computed by assigning to each cluster a weight equal
to the inverse value of the area for its {\it observed\/} flux. Such
area-flux relation is determined by the wavelet detection algorithm and is
therefore known precisely. This weight is further added to the flux bin in
correspondence to the {\it total\/} flux of the cluster.

While the calculation of the survey area as a function of cluster flux in
the detection cell is exact, the survey area as a function of total cluster
flux is known only approximately and exhibits a scatter, as illustrated in
the right panel of Fig.~\ref{f:logn}. Lower redshift systems require more
extrapolation for a given flux in the detection cell. Similarly, within the
same redshift range, more massive systems require larger degree of flux
extrapolation given the fixed size of the detection cell.

The left panel of Fig.~\ref{f:logn} shows the $\log(N>S)-\log(S)$ relation
of the COSMOS field clusters. Although with a somewhat higher normalization,
the COSMOS $\log(N>S)-\log(S)$ is statistically consistent with RDCS results
of Rosati et al. (2002) for the fluxes $S>10^{-14}$ ergs cm$^{-2}$
s$^{-1}$. While a similarly higher normalization of the $\log N - \log S$
has also been reported for the XMM-LSS survey (Pierre et al. 2005) as well
as for the 160 square degrees survey (Vikhlinin et al. 1998b), an important
difference between those surveys and the RDCS consists in extrapolation of
the cluster flux beyond the detection radius. As no extrapolation has been
done for RDCS, the difference in the results is due to higher flux being
assigned to each source, and not due to the higher source density. At fluxes
below $10^{-14}$ ergs cm$^{-2}$ s$^{-1}$ the XMM-COSMOS is the first survey
to yield rich observational data, allowing us to determine the
$\log(N>S)-\log(S)$ with good statistics down to $S\sim10^{-15}$ ergs
cm$^{-2}$ s$^{-1}$. We note that the prediction for no evolution in the
luminosity function obtained by local surveys, as summarized in Rosati et
al. (2002), provides a good fit to our cluster counts.

To check the uncertainties in our derived $\log(N>S)-\log(S)$ relation due
to our procedure for the estimate of the total flux, in Fig.~\ref{f:logn} we
show also the counts obtained if we compute the total flux for each clusters
assuming $\beta=0.6$, $r_{\rm core}=0.08r_{500}$ and using the parameters of
the $M-T$ relation from Vikhlinin et al. (2006). The number counts derived
with these assumptions are close to the upper envelope of the 1 sigma
confidence region of our $\log(N>S)-\log(S)$. This difference in
normalization is due to the fact that the $M-T$ of Vikhlinin implies a
larger size of the clusters, in particular at the lower mass end, compared
to the relation adopted here. We would like to point out several caveats in
the use of the presented $\log(N>S)-\log(S)$. One is that cluster
identification at $z>1.3$ has not been included, which can increase slightly
the faint end counts. Another caveat consists in the application of the
$\log(N>S)-\log(S)$ for predicting the number of clusters in surveys, as it
should take into account both the relation between the total flux and the
flux of the cluster in the detection cell, which is sensitive to the adopted
detection method. Finally, removal of the cool cores from the flux
estimates, results in an underestimate of the total X-ray flux from
clusters, typically by 20\% based on cluster studies at intermediate
redshifts (e.g. Zhang et al. 2006). A study of influence of cool cores on
both $\log(N>S)-\log(S)$ and the evolution of X-ray luminosity function will
be carried out within the granted Chandra program.

\subsection{Sample characteristics}

In Fig.~\ref{f:lxcat} we plot the observed characteristics of the XMM-COSMOS
cluster sample together with detection limits implied by both survey depth
and our approach to search for clusters of galaxies. The solid gray line has
been calculated by requiring that the $2\times r_{500}=32^{\prime\prime}$
and shows which sources cannot be detected as extended by our technique. The
dotted grey line is calculated imposing a criterion $2\times
r_{core}=32^{\prime\prime}$ and shows for which clusters the cores could be
resolved in our method. The dashed grey line is calculated by requiring that
$r_{500}=32^{\prime\prime}$ to reveal clusters for which removal of the
embedded point sources will result in a strong underestimate of the cluster
emission. The black lines show the detection limits of the survey achieved
over 90, 50 and 10\% of the total area. This comparison shows that with a
given method it is possible to go a factor of 10 deeper without losing X-ray
groups of galaxies that would appear point-like, which explains the success
of the application of our cluster detection method to deep fields. Thus, a
use of $32^{\prime\prime}$ scale does not introduce a systematics towards
the source detection, as for the flux limits of the survey the expected
extent of the X-ray emission is larger. On the other hand, resolving the
cores for the detected groups and clusters is difficult and requires a more
refined PSF modeling, that we postpone to a future effort. In all cases, the
central $0.5r_{500}$ is resolved. An underestimate of mass due to
oversubtraction of the unresolved cluster emission would occur between the
long-dashed and solid grey lines, and is currently below the survey
sensitivity limits.  We have already demonstrated that clusters with
$r_{500}$ smaller than our detection cell cannot be accessed based on the
sensitivity limits and do not introduce any systematics in the
selection. Likewise, the clusters much larger than the detection cell, which
are the brightest at each redshift, are all detected de facto, because the
available statistics allows us to map the core of the cluster at
$32^{\prime\prime}$ resolution. In a much shallower survey, where the
detection limits would be similar to the limiting flux at which the cluster
cores are resolved (e.g. for XMM it is the short-dashed curve in Fig.5), a
problem with a fixed detection size might occur.  Thus we conclude, that the
success of the method is a result of a good match between the detection
cell, the depth of the survey and the properties of X-ray emission of
clusters of galaxies.

\includegraphics[width=8.cm]{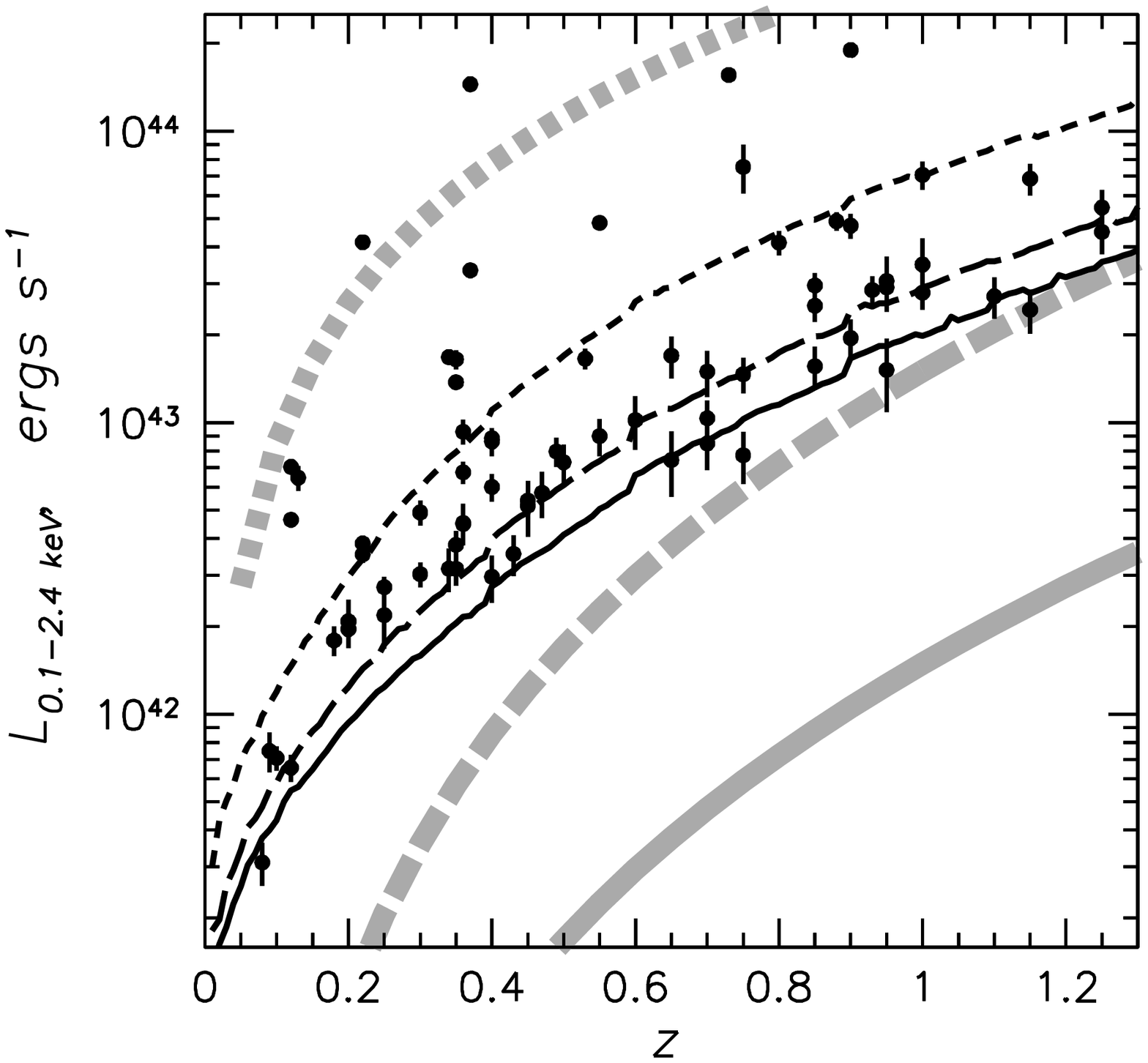}

\figcaption{Illustration of the cluster luminosity probed as a function of
cluster redshift in the XMM COSMOS survey. Filled circles represent the
detected clusters with error bars based on the statistical errors in the
flux measurements only. Short-dashed, long-dashed and solid black lines show
the flux detection limits associated with 90, 50 and 10\% of the total area,
respectively. Grey lines indicate the limits imposed by the detection method
(i.e. due to the angular resolution of the XMM-Newton telescopes). Systems
below the short-dashed grey line have unresolved cores, systems below the
long-dashed grey line (none in our sample) suffer from oversubtraction of
core emission. No system below the solid grey line can be detected as an
extended source using the method described in this paper.
\label{f:lxcat}}

In Fig.~\ref{f:dndz} we report the redshift distribution of both galaxy
overdensities and the identified X-ray structures. Galaxy overdensities are
determined from the analysis described in \S\ref{sec-cat}. In particular, we
display the histograms for the wavelet detected peaks in the photo-z slices
with $\Delta z=0.2$, which is also used to normalize the abundance of galaxy
structures in Fig.~\ref{f:dndz}.

Comparison between all galaxy overdensities with the X-ray selected systems
shows that 25\% of the galaxy systems are identified in X-rays at
$z<0.5$. At higher redshifts this proportion drops to 15\%. The reason for a
change in the identification rate consists in the increase of the total mass
of a group of galaxies that could be detected through its X-ray emission in
the XMM-COSMOS survey. At the same time, the depth of the catalog allows us
to cover the $L^*$ galaxies even at $z\simeq1.2$ and the abundance of galaxy
counterparts is not limiting the identification process at $z<1.3$.

\includegraphics[width=8.cm]{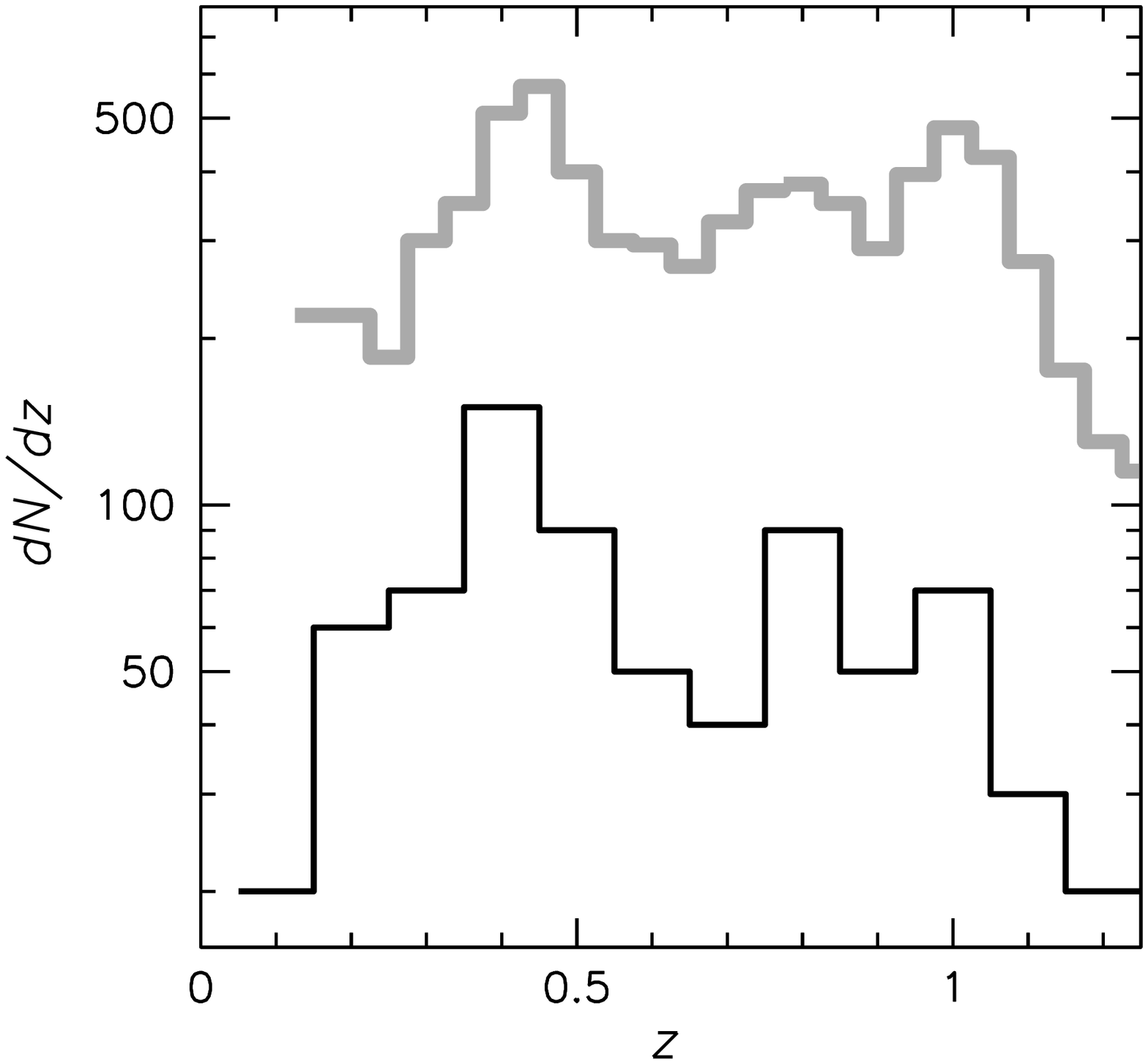}
\figcaption{Differential redshift distribution ($dN/dz$) of 420 photo-z galaxy
concentrations (grey histogram) as well as 72 X-ray (black histogram)
clusters of galaxies in the COSMOS field.
\label{f:dndz}}

The depth of the X-ray survey could also be characterized by the median
redshift of the catalog, which is shown in Fig.~\ref{f:cum}. The median
redshift for the total sample is 0.45. We define a subsample of the 12 most
massive objects in the sample: the galaxy clusters, which are traditionally
defined as $M_{200}>10^{14}M_\odot$, where $M_{200}\simeq1.7 M_{500}$. This
subsample is characterized by a median redshift of 0.75 and should not be
affected by our limiting X-ray flux, yet this value could be somewhat
underestimated, due to our $z<1.3$ limitation. Both the surface density of
clusters and the median redshift matches well the expectations for the
future SZ surveys. Thus, COSMOS field can be used to calibrate the
efficiency of SZ surveys.

\includegraphics[width=8.cm]{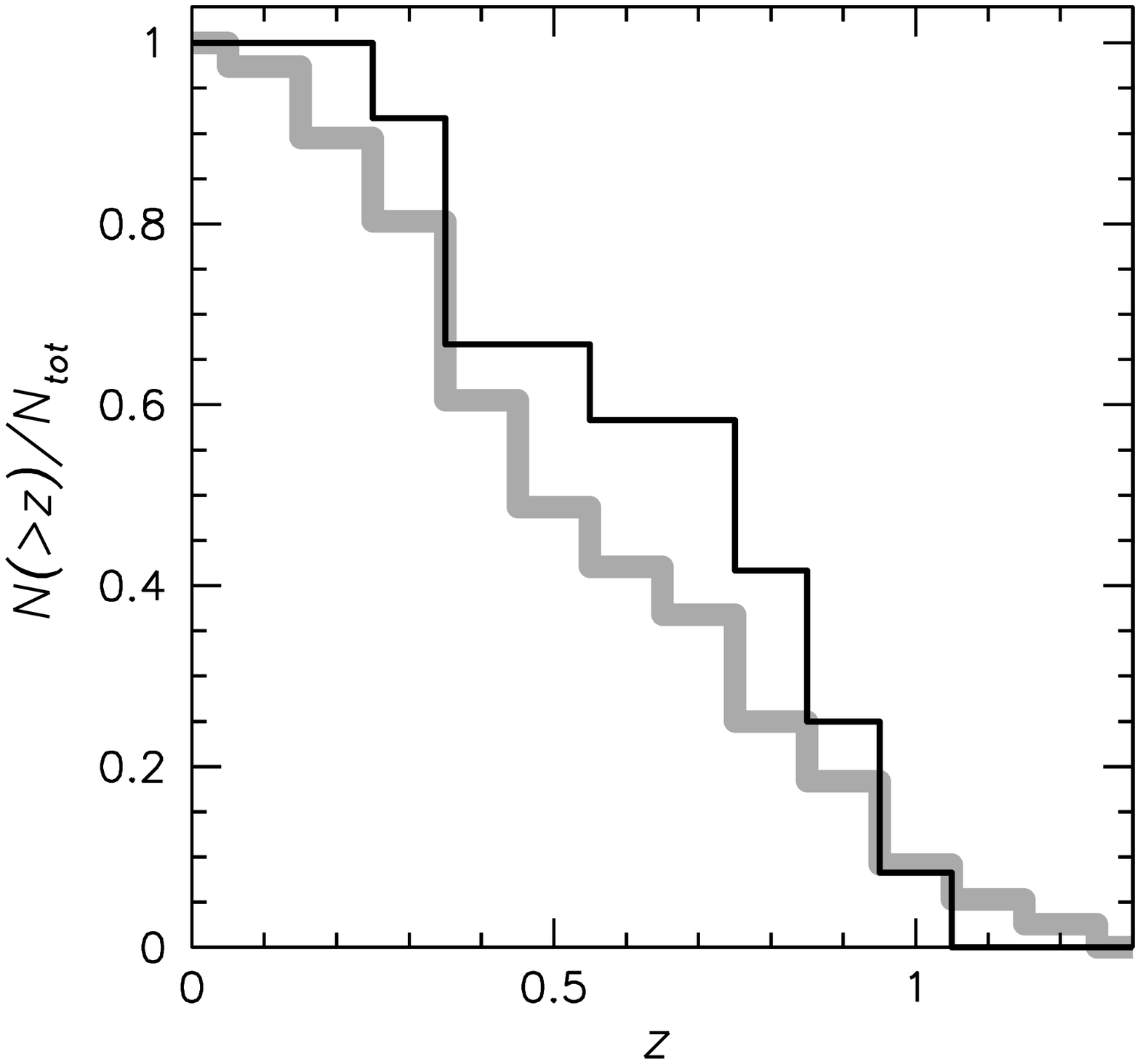} 

\figcaption{Normalized cumulative ($N(>z)/N_{\rm total}$) redshift
distribution of the diffuse X-ray sources, identified with a galaxy
concentration based on the photo-z's. The thick grey line represents the
full sample of 72 clusters, and the solid black line represents the 12
clusters ($M_{200}>10^{14}M_\odot$).
\label{f:cum}}

\section{X-ray luminosity function}\label{xfun}

The volume probed by the COSMOS survey is representative for LSS studies
(Scoville et al. 2007a) and the size of the COSMOS X-ray cluster catalog is
similar to many cluster surveys, which is due to a combination of high
sensitivity achieved over a sufficiently large area. For comparison, the
volume probed by the XMM-COSMOS survey to a redshift of 1.3 is achieved by
REFLEX (Boehringer et al. 2001) by $z\simeq0.05$, the 400\sq\deg\ survey
(Burenin et al. in prep.) by $z\simeq0.2$, and RDCS (Rosati et al. 2002) by
$z\simeq0.4$. At these redshifts, all those surveys provide volume-limited
samples of clusters brighter than a few times $10^{43}$ ergs s$^{-1}$. For
more luminous systems, wider surveys provide more volume, so the strength of
XMM-COSMOS consists in probing the redshift evolution of $10^{43}$ ergs
s$^{-1}$ clusters. To provide a more detailed comparison we construct here a
luminosity function for the clusters in our sample. The new flux regime
probed by the COSMOS survey yields an abundance of low-mass groups, so our
results will contribute to the refinement in the faint end of the luminosity
function.

In defining the luminosity function we followed the approach described in
Mullis et al. (2004). In particular, we used their Eq.4, which is adopted
from Page \& Carrera (2000) on the refined estimate of the luminosity
function. In calculating the maximum volume probed by the survey as a
function of cluster luminosity, we first tabulate the limiting luminosities
on the grid defined by tabulation of the survey area vs observed flux and
sample the redshifts 0--1.3 with a step of 0.01. We take into account the
typical extent to which the source is detected, which is found to be a
function of the observed flux, and estimate the degree of extrapolation
performed ($C_\beta(z,T)$) in obtaining the total flux and apply the
K-correction ($K(z,T)$).

In the no-evolution case and in the absence of strong clustering, one
expects to detect the clusters uniformly throughout the probed volume ($
\left< V/V_{max}\right> =0.5 $, Schmidt 1969). So, if any issues of
incomplete identification, e.g. resulting from the use of the photo-z
catalog, are important at some redshift (e.g. at very low redshifts or at
very high redshifts), this would introduce a distortion in the distribution
of clusters over the volume. To check if there are biases for some class of
objects (e.g. low-luminosity objects), we plot in Fig.~\ref{f:vmax} the
ratio between the volume towards the system and the maximum volume at which
it could be detected. As some values exceed unity due to the scatter in the
flux-area relation, we replaced them by unity for illustrative purposes. The
numbers of clusters above and below 0.5 are roughly equal at all
luminosities, indicating no large selection effects. The mean value of
$V/V_{max}$ for the survey is equal to $0.48\pm0.06$, which is consistent
with 0.5 within the statistical errors. This is an important result in
itself, that may illustrate that morphological changes observed in
high-redshift clusters (Postman et al. 2005) do not cause strong
redshift-dependent selection effects. It is clear, however, that any effects
of incompleteness occurring at the 10\% level would be hard to detect with
the size of our sample.

\includegraphics[width=8.cm]{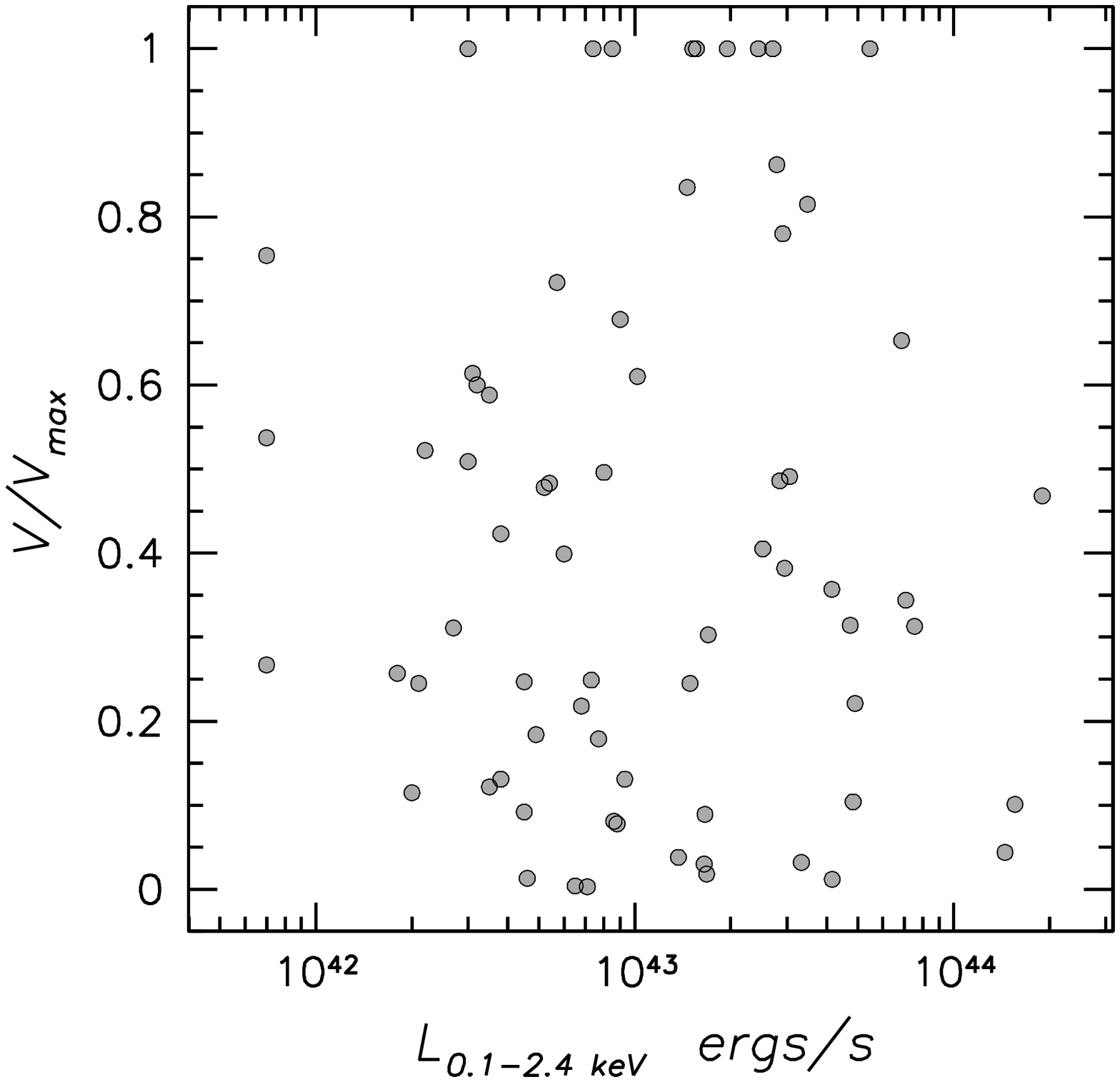} 
\figcaption{Test for the sample redshift completeness (V/V$_{\rm max}$). The
estimates exceeding 1 (due to the scatter in the flux-area relation) are
substituted with 1.  The number of clusters above and below 0.5 is roughly
equal at all luminosities, indicating no large selection effects.
\label{f:vmax}}

Finally, in Fig.~\ref{f:lxfunc} we present the luminosity function of XMM
COSMOS clusters. A comparison with the results of the REFLEX (Boehringer et
al. 2001) and the BCS survey (Ebeling et al. 1997) displayed in
Fig.~\ref{f:lxfunc} shows that, at the luminosity range probed by the COSMOS
survey, the evolution in the luminosity function is not statistically
significant.

The COSMOS data allow us to put tight constraints on the slope of the faint
end of the luminosity function. To characterize it, we fitted a Schechter
function to the data adopting the $L_X^*$ and $\phi^*$ parameters in
correspondence to the best fit values of BCS survey ($9.1\times10^{44}
h_{50}^{-2}$ ergs/s$^{-1}$ and $7.74\times10^{-8} h_{50}^3$ Mpc$^{-3}$). We
achieve an acceptable value of the reduced $\chi^2=0.7$ for 7 degrees of
freedom and constrain the value of the slope to $\alpha=1.93\pm0.04$ (where
$dn/dL \propto L^{-\alpha}$), using the Gehrels (1986) approximations in
calculating the confidence limits for the case of small number
statistics. The value of the slope compares well with the BCS result of
$\alpha=1.85\pm0.09$. Our slope value is also within the uncertainty
reported for the 160 square degrees survey (Mullis et al. 2004). However,
their use of the $0.5-2$ keV energy band results in a somewhat lower value
of the slope, and a strict comparison of luminosity functions is
difficult. The lower number of groups in the REFLEX survey (dotted line in
Fig.~\ref{f:lxfunc}) is thought to be due to a combination of the small
survey depth at low luminosities and a presence of a local southern void,
where most of the survey area is located and whose effect has been
demonstrated through differences within the sample (B\"ohringer et
al. 2002).

To illustrate the lack of redshift evolution in the luminosity function, in
Fig.~\ref{f:lxfunc} we split the sample in two redshift bins, 0--0.6 (dotted
crosses) and 0.6--1.3 (solid crosses), retaining only the luminosity bins
derived using at least three clusters. The two subsamples overlap only in a
single luminosity bin, where the corresponding cluster abundances agree
within the errors.  Since the high-luminosities are well probed only at
redshifts higher than 0.6, the good match between our measurements and the
local luminosity function is an indication of the absence of a significant
redshift evolution. This finding is in agreement with the results of Mullis
et al. (2004), where detectable evolutionary effects are seen just above
$L_x\sim10^{44}$ ergs/s. We note that our measurements are of comparable
quality to the existing data compiled in Mullis et al. (2004) and provide a
refinement to the knowledge on the luminosity function at high redshifts. A
further refinement of the COSMOS survey stems from its ability to measure
the mass independently via weak lensing and therefore reduce the effects of
systematics in determining the mass function. The study of cluster evolution
provides competitive constraints in the $\Omega_{M}-\Omega_{\Lambda}$ plane,
and yields results in agreement with other measurements (Vikhlinin
et al.  2003). Finally, we note that both our cluster counts and the
luminosity function are consistent with no evolution in the luminosity
function in the $8\times10^{42}-2\times10^{44}$ ergs s$^{-1}$ range. This
provides further evidence in favor of a consistent modeling of the
XMM-COSMOS survey sensitivity presented in this paper.

\includegraphics[width=8.cm]{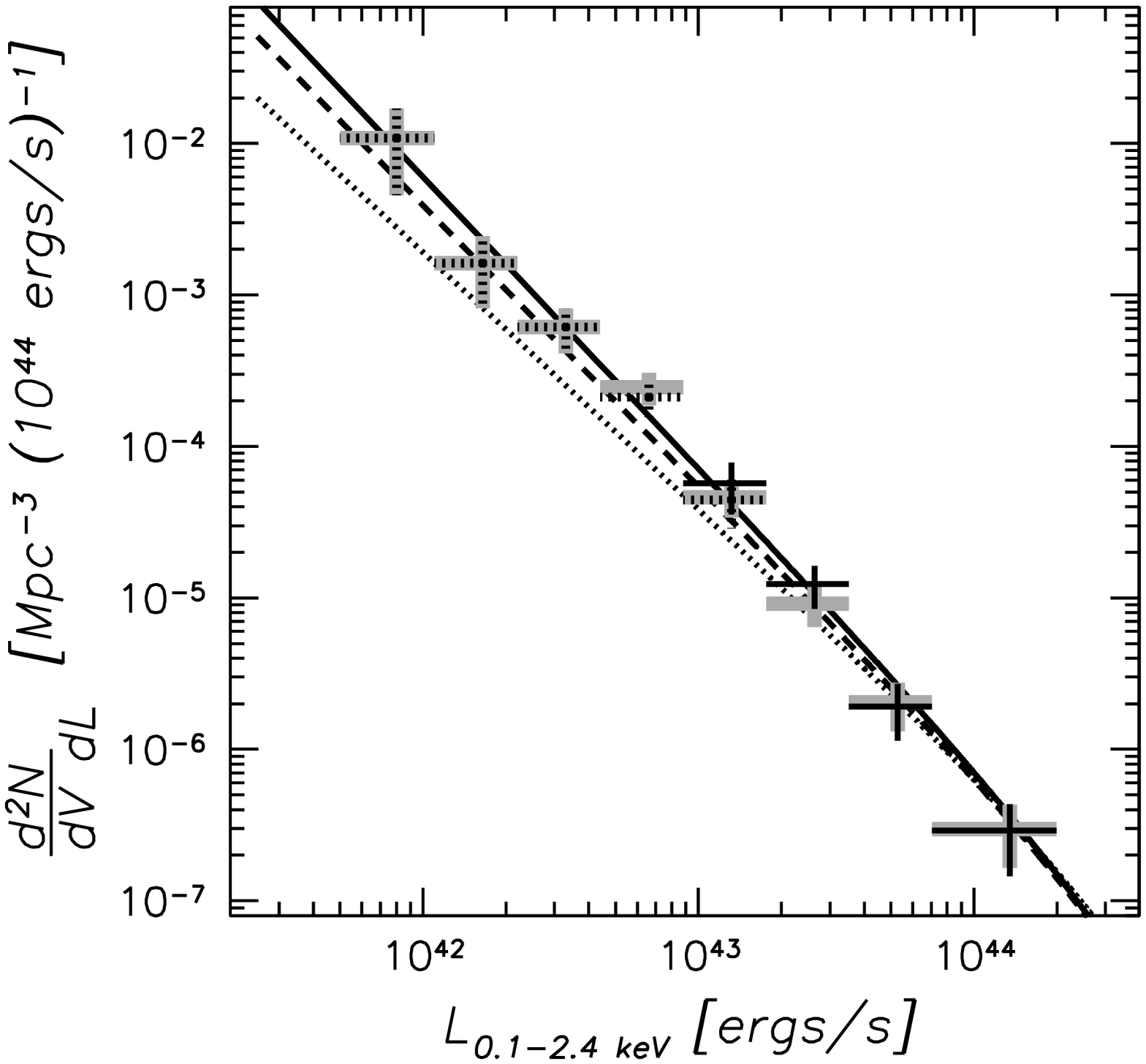}

\figcaption{Luminosity function of clusters in the COSMOS field. Dotted
crosses indicate the data in the redshift range 0--0.6, grey points are the
data in the redshift range 0--1.3 and solid crosses indicate the data in the
redshift range 0.6--1.3. The dotted line shows the luminosity function of
the REFLEX survey ($0<z<0.3$, B\"ohringer et al.2001) and the dashed line
shows the results of BCS survey (Ebeling et al. 1997), which illustrates the
current uncertainly on the shape of the luminosity function at $z<0.3$. The
solid line shows the best fit to the COSMOS data.
\label{f:lxfunc}}

\section{Summary}\label{resume}

We present a description of our X-ray based cluster detection method and the
first results of the cluster search using the XMM-COSMOS survey. Our flux
range is $3\times10^{-15}-10^{-13}$ erg cm$^{-2}$ s$^{-1}$ in the 0.5--2 keV
band. We run a separate analysis of the photo-z catalog to identify 420
early-type galaxy concentrations, which provide an identification to 72
X-ray cluster candidates. We further present the statistics for those
clusters in terms of $\log N-\log S$, $dN/dz$ and $dn/dL$. By comparison
with local cluster surveys, we find no evolution in cluster number abundance
out to a redshift of 1.3 in the luminosity range of $L_{0.1-2.4 \rm keV}:
8\times10^{42}-2\times10^{44}$ ergs s$^{-1}$. This further implies that the
surface density of clusters detected in the flux range $10^{-15}-10^{-14}$
erg cm$^{-2}$ s$^{-1}$ should correspond to the prediction of no evolution,
higher than implied by Rosati et al. (2002). Such high surface density of
clusters has been found by both COSMOS and XMM-LSS surveys (Pierre et
al. 2005). The published results on deeper X-ray surveys, CDFS and CDFN,
contradict this view (Rosati et al. 2002). It is therefore important to
understand the origin of this inconsistency, which is likely related to
evolution of cluster X-ray luminosity function at $z>1.2$, probed at fluxes
fainter than $10^{-15}$ erg cm$^{-2}$ s$^{-1}$.

X-ray selected clusters provide well-defined information on the most massive
high density peaks in the COSMOS field and allow a follow-up study of the
evolution in the galaxy morphology in dense environments.
 
\acknowledgments

In Germany, the XMM--Newton project is supported by the Bundesministerium
fuer Wirtschaft und Technologie/Deutsches Zentrum fuer Luft- und Raumfahrt
(BMWI/DLR, FKZ 50 OX 0001), the Max-Planck Society and the
Heidenhain-Stiftung. Part of this work was supported by the Deutsches
Zentrum f\"ur Luft-- und Raumfahrt, DLR project numbers 50 OR 0207 and 50 OR
0405.  The HST COSMOS Treasury program was supported through NASA grant
HST-GO-09822. We gratefully acknowledge the contributions of the entire
COSMOS collaboration consisting of more than 70 scientists. More information
on the COSMOS survey is available at {\tt
  \url{http://www.astro.caltech.edu/cosmos}}. It is a pleasure to
acknowledge the excellent services provided by the NASA IPAC/IRSA staff
(Anastasia Laity, Anastasia Alexov, Bruce Berriman and John Good) in
providing online archive and server capabilities for the COSMOS datasets.
The COSMOS Science meeting in May 2005 was supported in part by the NSF
through grant OISE-0456439. AF acknowledges support from BMBF/DLR under
grant 50 OR 0207, MPG and a partial support from NASA grant NNG04GF686,
covering his stays at UMBC. The authors thank the referee, Alastair Edge,
for detailed comments, which improved the content of this paper. AF thanks
Harald Ebeling, Trevor Ponman, Piero Rosati, Peter Schuecker for useful
suggestions at various stages of this work.  Irini Sakelliou acknowledges
the support of the European Community under a Marie Curie Intra-European
Fellowship.


 \end{document}

%% file: cat_tab4.tex
\begin{deluxetable}{lccccccccccc}
\tablewidth{0pt}
\tabletypesize{\footnotesize}
\tablecaption{Catalog of the identified X-ray clusters.}
\tablehead{
\colhead{ } &
\colhead{R.A} &  
\colhead{Decl.} & 
\colhead{$r_{500}$} & 
\colhead{flux  $10^{-14}$} &
\colhead{z} &
\colhead{M$_{500}$} & 
\colhead{L$_{\rm 0.1-2.4 keV}$} & 
\colhead{kT} &
\colhead{z}\\
\colhead{ID} & 
\multicolumn{2}{c}{Eq.2000} & 
\colhead{$\prime$} &
\colhead{ergs cm$^{-2}$ s$^{-1}$} & 
\colhead{ } & 
\colhead{$10^{13}$ M$_\odot$} & 
\colhead{$10^{42}$ ergs s$^{-1}$} & 
\colhead{keV}  & 
\colhead{source} 
}
\startdata
  3 & 150.80244 & 1.98985 & 1.0 & $ 0.70\pm0.16$ & 0.25 & $ 0.69\pm0.14$ &  $2.2\pm 0.5$ & $0.56\pm0.06$& 1\\  
  9 & 150.75121 & 1.52793 & 1.1 & $ 1.97\pm0.38$ & 0.75 & $ 9.62\pm1.65$ &$ 75.4\pm14.5$ & $2.62\pm0.23$& 1\\
 11 & 150.73676 & 2.82680 & 0.7 & $ 0.40\pm0.08$ & 0.60 & $ 1.88\pm0.35$ &$ 10.2\pm 2.1$ & $1.06\pm0.10$& 1\\
 15 & 150.67342 & 2.09190 & 0.9 & $ 0.50\pm0.09$ & 0.34 & $ 0.88\pm0.13$ &$  3.2\pm 0.5$ & $0.65\pm0.05$& 0\\
 20 & 150.64041 & 2.12791 & 0.8 & $ 0.43\pm0.06$ & 0.55 & $ 1.78\pm0.23$ &$  9.0\pm 1.3$ & $1.01\pm0.07$& 1\\
 24 & 150.58962 & 2.87187 & 0.7 & $ 0.36\pm0.08$ & 0.95 & $ 3.41\pm0.66$ &$ 30.6\pm 6.6$ & $1.61\pm0.16$& 1\\
 25 & 150.58631 & 1.92693 & 1.1 & $ 1.04\pm0.10$ & 0.30 & $ 1.36\pm0.12$ &$  4.9\pm 0.5$ & $0.81\pm0.04$& 1\\
 29 & 150.53842 & 2.37393 & 1.3 & $ 1.28\pm0.15$ & 0.18 & $ 0.62\pm0.07$ &$  1.8\pm 0.2$ & $0.52\pm0.03$& 0\\
 32 & 150.50535 & 2.22395 & 1.2 & $ 3.45\pm0.11$ & 0.90 & $18.64\pm0.52$ &$189.7\pm 5.9$ & $3.90\pm0.06$& 1\\
 34 & 150.49330 & 2.06795 & 0.9 & $ 0.61\pm0.07$ & 0.40 & $ 1.47\pm0.14$ &$  6.0\pm 0.6$ & $0.87\pm0.04$& 1\\
 36 & 150.49048 & 2.74592 & 0.6 & $ 0.22\pm0.05$ & 0.65 & $ 1.34\pm0.30$ &$  7.4\pm 1.9$ & $0.90\pm0.10$& 1\\
 38 & 150.44824 & 1.91197 & 0.6 & $ 0.26\pm0.04$ & 1.25 & $ 3.49\pm0.50$ &$ 45.1\pm 7.3$ & $1.79\pm0.13$& 1\\
 39 & 150.44827 & 2.04996 & 1.3 & $ 2.54\pm0.13$ & 0.55 & $ 8.10\pm0.38$ &$ 48.4\pm 2.5$ & $2.25\pm0.05$& 1\\
 40 & 150.44523 & 1.88197 & 0.6 & $ 0.26\pm0.04$ & 0.70 & $ 1.70\pm0.23$ &$ 10.4\pm 1.6$ & $1.03\pm0.07$& 1\\
 41 & 150.44229 & 2.15796 & 0.7 & $ 0.30\pm0.06$ & 0.40 & $ 0.77\pm0.13$ &$  3.0\pm 0.6$ & $0.62\pm0.05$& 1\\
 42 & 150.41533 & 2.43096 & 2.8 & $11.45\pm0.26$ & 0.12 & $ 2.26\pm0.05$ &$  7.1\pm 0.2$ & $1.01\pm0.01$& 0\\
 44 & 150.42124 & 1.98397 & 0.8 & $ 0.42\pm0.07$ & 0.45 & $ 1.26\pm0.19$ &$  5.4\pm 0.9$ & $0.81\pm0.06$& 1\\
 45 & 150.42121 & 1.84898 & 0.7 & $ 0.49\pm0.05$ & 0.85 & $ 3.70\pm0.35$ &$ 29.5\pm 3.1$ & $1.63\pm0.08$& 1\\
 47 & 150.40934 & 2.51196 & 0.6 & $ 0.29\pm0.04$ & 1.00 & $ 2.97\pm0.34$ &$ 27.9\pm 3.5$ & $1.52\pm0.09$& 1\\
 51 & 150.37616 & 1.66900 & 0.7 & $ 0.32\pm0.04$ & 0.75 & $ 2.20\pm0.28$ &$ 14.6\pm 2.1$ & $1.20\pm0.08$& 1\\
 52 & 150.37930 & 2.40997 & 0.8 & $ 0.47\pm0.06$ & 0.35 & $ 0.87\pm0.10$ &$  3.1\pm 0.4$ & $0.65\pm0.04$& 0\\
 53 & 150.37021 & 1.99898 & 0.7 & $ 0.41\pm0.05$ & 0.85 & $ 3.20\pm0.34$ &$ 25.2\pm 3.0$ & $1.51\pm0.08$& 1\\
 54 & 150.33413 & 1.60301 & 1.0 & $ 0.88\pm0.10$ & 0.40 & $ 2.03\pm0.20$ &$  8.6\pm 0.9$ & $1.03\pm0.05$& 1\\
 56 & 150.31318 & 2.00799 & 0.9 & $ 0.57\pm0.07$ & 0.35 & $ 1.03\pm0.11$ &$  3.8\pm 0.4$ & $0.71\pm0.04$& 1\\
 57 & 150.28611 & 1.55502 & 1.1 & $ 1.27\pm0.13$ & 0.36 & $ 2.26\pm0.20$ &$  9.3\pm 0.9$ & $1.08\pm0.05$& 0\\
 58 & 150.28916 & 2.28024 & 1.4 & $ 0.91\pm0.10$ & 0.12 & $ 0.27\pm0.03$ &$  0.7\pm 0.1$ & $0.32\pm0.02$& 0\\
 59 & 150.28311 & 1.57902 & 0.9 & $ 0.62\pm0.10$ & 0.36 & $ 1.18\pm0.17$ &$  4.5\pm 0.7$ & $0.77\pm0.06$& 0\\
 62 & 150.21114 & 2.28100 & 0.8 & $ 0.79\pm0.06$ & 0.88 & $ 5.65\pm0.37$ &$ 49.1\pm 3.6$ & $2.06\pm0.07$& 0\\
 64 & 150.23218 & 2.48199 & 0.9 & $ 0.64\pm0.06$ & 0.30 & $ 0.88\pm0.08$ &$  3.0\pm 0.3$ & $0.64\pm0.03$& 1\\
 65 & 150.21111 & 1.81600 & 0.9 & $ 0.89\pm0.07$ & 0.53 & $ 3.16\pm0.23$ &$ 16.6\pm 1.4$ & $1.36\pm0.05$& 0\\
 66 & 150.21718 & 2.73998 & 0.5 & $ 0.15\pm0.04$ & 0.95 & $ 1.81\pm0.45$ &$ 15.2\pm 4.3$ & $1.16\pm0.15$& 1\\
 67 & 150.19609 & 1.65701 & 2.7 & $17.94\pm0.30$ & 0.22 & $10.11\pm0.15$ &$ 41.6\pm 0.7$ & $2.29\pm0.02$& 0\\
 68 & 150.21115 & 2.40100 & 0.6 & $ 0.24\pm0.04$ & 0.90 & $ 2.40\pm0.34$ &$ 19.5\pm 3.1$ & $1.32\pm0.10$& 1\\
 70 & 150.18109 & 1.76801 & 1.3 & $ 2.09\pm0.11$ & 0.35 & $ 3.27\pm0.15$ &$ 13.7\pm 0.7$ & $1.31\pm0.03$& 0\\
 71 & 150.19616 & 2.82397 & 1.2 & $ 1.14\pm0.21$ & 0.20 & $ 0.70\pm0.12$ &$  2.1\pm 0.4$ & $0.55\pm0.05$& 1\\
 72 & 150.16011 & 2.60499 & 0.8 & $ 0.69\pm0.07$ & 0.90 & $ 5.34\pm0.47$ &$ 47.4\pm 4.6$ & $2.02\pm0.09$& 1\\
 73 & 150.16912 & 2.52400 & 0.5 & $ 0.15\pm0.03$ & 0.75 & $ 1.24\pm0.22$ &$  7.7\pm 1.6$ & $0.89\pm0.08$& 1\\
 75 & 150.15410 & 2.39500 & 0.5 & $ 0.16\pm0.03$ & 1.15 & $ 2.23\pm0.34$ &$ 24.4\pm 4.2$ & $1.37\pm0.11$& 1\\
 78 & 150.11807 & 2.35600 & 1.3 & $ 1.74\pm0.11$ & 0.22 & $ 1.19\pm0.07$ &$  3.8\pm 0.2$ & $0.74\pm0.02$& 0\\
 79 & 150.11807 & 2.68299 & 1.4 & $ 2.46\pm0.19$ & 0.35 & $ 3.84\pm0.26$ &$ 16.5\pm 1.2$ & $1.42\pm0.05$& 0\\
 80 & 150.10906 & 2.55700 & 0.8 & $ 0.44\pm0.07$ & 0.50 & $ 1.56\pm0.21$ &$  7.3\pm 1.1$ & $0.93\pm0.06$& 1\\
 82 & 150.10606 & 2.42200 & 1.0 & $ 0.76\pm0.07$ & 0.22 & $ 0.59\pm0.05$ &$  1.8\pm 0.2$ & $0.51\pm0.02$& 0\\
 83 & 150.10906 & 2.01400 & 0.6 & $ 0.25\pm0.04$ & 0.85 & $ 2.09\pm0.31$ &$ 15.6\pm 2.6$ & $1.21\pm0.09$& 0\\
 84 & 150.09405 & 2.20000 & 0.7 & $ 0.36\pm0.04$ & 0.93 & $ 3.29\pm0.34$ &$ 28.5\pm 3.3$ & $1.57\pm0.08$& 0\\
 85 & 150.09105 & 2.39500 & 1.3 & $ 1.58\pm0.10$ & 0.22 & $ 1.10\pm0.06$ &$  3.5\pm 0.2$ & $0.71\pm0.02$& 0\\
 86 & 150.09705 & 2.30200 & 0.9 & $ 0.63\pm0.06$ & 0.36 & $ 1.18\pm0.10$ &$  4.5\pm 0.4$ & $0.76\pm0.04$& 0\\
 87 & 150.05802 & 2.38000 & 1.0 & $ 0.91\pm0.08$ & 0.40 & $ 2.08\pm0.16$ &$  8.8\pm 0.7$ & $1.04\pm0.04$& 1\\
 89 & 150.03999 & 2.69499 & 1.2 & $ 1.04\pm0.14$ & 0.20 & $ 0.66\pm0.08$ &$  2.0\pm 0.3$ & $0.54\pm0.03$& 1\\
 93 & 150.04300 & 2.54500 & 0.6 & $ 0.34\pm0.05$ & 1.25 & $ 4.15\pm0.56$ &$ 54.6\pm 8.2$ & $1.96\pm0.13$& 1\\
 97 & 149.98594 & 2.58099 & 0.6 & $ 0.21\pm0.04$ & 0.70 & $ 1.43\pm0.24$ &$  8.5\pm 1.6$ & $0.94\pm0.08$& 1\\
 99 & 149.96500 & 1.68101 & 1.6 & $ 4.38\pm0.26$ & 0.37 & $ 7.05\pm0.38$ &$ 33.3\pm 2.0$ & $1.98\pm0.06$& 0\\
100 & 149.97091 & 2.78197 & 0.7 & $ 0.39\pm0.07$ & 0.70 & $ 2.37\pm0.39$ &$ 15.0\pm 2.7$ & $1.23\pm0.10$& 1\\
101 & 149.96495 & 2.21199 & 0.7 & $ 0.31\pm0.05$ & 0.43 & $ 0.89\pm0.13$ &$  3.5\pm 0.6$ & $0.67\pm0.05$& 0\\
102 & 149.95293 & 2.34099 & 0.5 & $ 0.20\pm0.03$ & 1.10 & $ 2.60\pm0.38$ &$ 27.1\pm 4.4$ & $1.46\pm0.11$& 1\\
103 & 149.94991 & 2.48199 & 0.8 & $ 0.84\pm0.08$ & 0.80 & $ 5.31\pm0.46$ &$ 41.5\pm 4.0$ & $1.95\pm0.09$& 1\\
104 & 149.94987 & 2.91995 & 2.7 & $ 9.96\pm0.98$ & 0.13 & $ 2.08\pm0.18$ &$  6.5\pm 0.6$ & $0.97\pm0.04$& 0\\
105 & 149.91987 & 2.60198 & 1.1 & $ 0.88\pm0.08$ & 0.25 & $ 0.84\pm0.07$ &$  2.7\pm 0.2$ & $0.62\pm0.03$& 1\\
106 & 149.91688 & 2.51498 & 1.4 & $ 4.56\pm0.13$ & 0.73 & $18.90\pm0.48$ &$155.7\pm 4.4$ & $3.73\pm0.05$& 0\\
108 & 149.88981 & 2.80596 & 0.8 & $ 0.53\pm0.09$ & 0.65 & $ 2.82\pm0.41$ &$ 17.0\pm 2.8$ & $1.33\pm0.10$& 1\\
111 & 149.88386 & 2.44898 & 1.0 & $ 0.96\pm0.09$ & 0.36 & $ 1.70\pm0.13$ &$  6.8\pm 0.6$ & $0.93\pm0.04$& 0\\
113 & 149.85995 & 1.76499 & 2.4 & $ 7.44\pm0.44$ & 0.12 & $ 1.55\pm0.08$ &$  4.6\pm 0.3$ & $0.83\pm0.02$& 0\\
114 & 149.81183 & 2.25397 & 0.8 & $ 0.39\pm0.07$ & 0.47 & $ 1.29\pm0.21$ &$  5.7\pm 1.0$ & $0.83\pm0.07$& 0\\
117 & 149.77272 & 2.63795 & 1.7 & $ 1.03\pm0.17$ & 0.08 & $ 0.14\pm0.02$ &$  0.3\pm 0.1$ & $0.23\pm0.02$& 0\\
119 & 149.74586 & 1.94797 & 0.8 & $ 0.41\pm0.09$ & 0.45 & $ 1.22\pm0.24$ &$  5.2\pm 1.1$ & $0.80\pm0.08$& 1\\
120 & 149.75467 & 2.79393 & 0.8 & $ 0.50\pm0.06$ & 0.49 & $ 1.70\pm0.18$ &$  8.0\pm 0.9$ & $0.97\pm0.05$& 0\\
126 & 149.64969 & 2.34093 & 0.8 & $ 0.87\pm0.10$ & 1.00 & $ 6.87\pm0.69$ &$ 70.8\pm 7.9$ & $2.37\pm0.12$& 1\\
128 & 149.64372 & 2.21193 & 0.6 & $ 0.38\pm0.09$ & 1.00 & $ 3.63\pm0.75$ &$ 34.8\pm 8.1$ & $1.69\pm0.18$& 1\\
132 & 149.59548 & 2.82087 & 1.4 & $ 2.57\pm0.13$ & 0.34 & $ 3.92\pm0.18$ &$ 16.8\pm 0.9$ & $1.44\pm0.04$& 1\\
133 & 149.60532 & 2.43541 & 0.7 & $ 0.57\pm0.07$ & 1.15 & $ 5.68\pm0.64$ &$ 68.7\pm 8.6$ & $2.25\pm0.13$& 1\\
134 & 149.60148 & 2.85087 & 0.6 & $ 0.34\pm0.05$ & 0.95 & $ 3.26\pm0.46$ &$ 29.1\pm 4.6$ & $1.58\pm0.11$& 1\\
140 & 149.48149 & 2.51785 & 1.9 & $ 2.04\pm0.32$ & 0.09 & $ 0.31\pm0.04$ &$  0.7\pm 0.1$ & $0.35\pm0.03$& 0\\
145 & 149.39739 & 2.57480 & 2.5 & $20.42\pm0.46$ & 0.37 & $26.47\pm0.54$ &$144.9\pm 3.3$ & $3.98\pm0.04$& 0\\
\enddata
\end{deluxetable}